\documentclass[twocolumn]{aastex631}

\usepackage{graphicx}
\usepackage{amsmath}
\usepackage{xcolor}
\usepackage{dcolumn}
\usepackage{colortbl}
\usepackage{booktabs}
\usepackage{bm}

\newcolumntype{?}{!{\vrule width 0.6 pt}}
\newcommand{\thickhline}{\noalign{\hrule height 0.7pt}}

\begin{document}

\title{
Uniting the Observed Dynamical Dark Energy Preference with the Discrepancies in $\Omega_m$ and $H_0$ Across Cosmological Probes
}

\author[0009-0007-3185-7030]{Xianzhe TZ Tang}
\email{tztang@bu.edu}
\affiliation{Boston University Department of Astronomy, 725 Commonwealth Ave, Boston USA}
\author[0000-0001-5201-8374]{Dillon Brout}
\email{dbrout@bu.edu}
\affiliation{Boston University Department of Astronomy, 725 Commonwealth Ave, Boston USA}
\affiliation{Boston University Department of Physics, 725 Commonwealth Ave, Boston USA}
\author[0000-0002-1384-9949]{Tanvi Karwal} 
\affiliation{Kavli Institute for Cosmological Physics, University of Chicago, IL 60637, USA}
\author[0000-0002-7887-0896]{Chihway Chang} 
\affiliation{Kavli Institute for Cosmological Physics, University of Chicago, IL 60637, USA}
\affiliation{Department of Astronomy and Astrophysics, University of Chicago, Chicago, IL 60637, USA}
\author[0000-0003-4776-0333]{Vivian Miranda}
\affiliation{C. N. Yang Institute for Theoretical Physics, Stony Brook University, Stony Brook, NY 11794, USA}
\author[0000-0001-8788-1688]{Maria Vincenzi} 
\affiliation{Department of Physics, Oxford University, Oxford, UK}

\date{\today}

\begin{abstract}

Recent results from Type Ia Supernovae (SNe), baryon acoustic oscillations (BAO), and the cosmic microwave background (CMB) indicate 1) potentially discrepant measurements of the matter density $\Omega_m$ and Hubble constant $ H_0 $ in $\Lambda$CDM model when analyzed individually, and 2) hints of dynamical dark energy in a $w_0w_a$CDM model when data are combined in a joint analysis. 
We examine whether underlying dynamical dark energy cosmologies favored by data would result in biases in $\Omega_m$ and $ H_0 $ for each probe when analyzed individually under $\Lambda$CDM. 
We generate mock datasets in $w_0w_a$CDM cosmologies, fit the individual probes under the $\Lambda$CDM model, and find expected biases in $\Omega_m$ are $\sim 0.03$. 
Notably, the $\Omega_m$ differences between probes are consistent with values observed in real datasets. 
We also observe that mock DESI-BAO datasets generated in the $w_0w_a$CDM cosmologies will lead to a biased measurement of $H_0$ higher by ($\sim1.2$km/s/Mpc) when fitted under $\Lambda$CDM, appearing to mildly improve the Hubble tension, but as the true underlying $H_0$ is lower, the tension is in fact worsened. We find that the $\Omega_m$ discrepancies, the high BAO $ H_0 $ relative to CMB, and the joint dynamical dark energy signal are all related effects that could be explained \textit{simultaneously} with either new physics or new systematics.
While it is possible to unite many of the discrepancies seen in recent analyses along a single axis, our results underscore the importance of understanding systematic differences in datasets, as they have unique impacts in different cosmological parameter spaces. 

\end{abstract}

\section{\label{sec:level1} Introduction}

The standard $\Lambda$CDM cosmological model, which includes a cosmological constant dark energy \(\Lambda\) and cold dark matter (CDM), has historically been remarkably successful in explaining a wide range of cosmological observations, from measurements of the cosmic microwave background (CMB) \citep{Planck2018} to Type Ia supernovae (SNe) \citep{brout2022pantheon+, DESY5}. 

However, recent measurements of baryon acoustic oscillations (BAO) from the Dark Energy Spectroscopic Instrument (DESI) and SNe from the Dark Energy Survey (DES) have revealed intriguing discrepancies in the value of \(\Omega_m\) under the \(\Lambda\)CDM model when compared to other analyses with CMB and SNe data. 
Specifically, BAO data from DESI Y1 tend to favor lower values of \(\Omega_m = 0.295 \pm 0.015\) \citep{desiy1}, and the new DESI full shape-modeling of clustering measurements with Big Bang Nucleosynthesis (BBN) and scalar spectral index priors find \(\Omega_m = 0.2962 \pm 0.0095\) \citep{DESIFS}.
On the other hand, SN datasets prefer higher values. 
For instance, 
\(\Omega_m = 0.353 \pm 0.017\) from DES-SN5YR \citep{DESY5,vincenzi24,sánchez2024darkenergysurveysupernova}, 
\(\Omega_m = 0.331 \pm 0.018\) from PantheonPlus \citep{brout2022pantheon+,Scolnic_2022}, and 
\(\Omega_m = 0.359\pm0.027\) from Union3 \citep{rubin2023union}. 
Although these differences do not reach the 3\(\sigma\) threshold for significant tension, they hint at possible inconsistencies among the datasets, with statistical deviations of 1.6\(\sigma\) for PantheonPlus, 2.0\(\sigma\) for Union3, and 2.6\(\sigma\) for DES-SN5YR when compared to DESI results \citep{desiy1}. 
Further quantifying this tension, \cite{berghaus2024quantifying} analyzed the combined CMB+DESI+DES-SN5YR data (CMB being Planck $C_{\ell}^{\rm TTTEEE}$ restricted to $\ell < 1296$) and observed a \(\Delta \chi^2\) of approximately 8.5 when \(\Omega_m = 0.295\), corresponding to the best-fit value from DESI BAO measurements. This chi-square difference translates to a tension of approximately 2.9 sigma, indicating a significant deviation in \(\Omega_m\) between the supernova datasets and the DESI BAO preferred cosmology.

In addition to discrepancies in \( \Omega_m \), recent observations have revealed differences in the Hubble constant (\( H_0 \)) between BAO and CMB measurements. 
The DESI Y1 BAO results \citep{desiy1}, when combined with calibrations of the sound horizon \( r_d \) from the CMB \citep{Planck2018} or BBN \citep{BBN}, yield higher \( H_0 \) values than the CMB itself. 
Specifically, DESI Y1 BAO find \( H_0 = 69.29 \pm 0.87 \) km/s/Mpc with \( r_d \) from the CMB or \( H_0 = 68.53 \pm 0.80 \) km/s/Mpc with a BBN prior \citep{desiy1}. 
Recently, the DESI 2024 VII full shape with BBN and $n_{s10}$ prior shows \( H_0 = 68.56 \pm 0.75 \) km/s/Mpc \citep{DESIFS}. 
These BAO values are notably higher than the \( H_0 = 67.27 \pm 0.60 \) km/s/Mpc (TT, TE, EE, LowE) reported by the Planck 2018 CMB analysis \citep{Planck2018} by about 1--2 km/s/Mpc.
While these differences in \( H_0 \) between BAO and CMB are not at the level of the Hubble Tension between Planck and SH0ES-calibrated SNe ($\sim6$km/s/Mpc at $>5\sigma$) \citep{riess2022comprehensive}, it is a notable difference that has withstood several recent analyses.

Moreover and perhaps most notably, recent analyses have also hinted at the possibility of evolving dark energy when all datasets (SNe, BAO, CMB) are combined \citep{brout2022pantheon+,DESY5,desiy1}. 
Specifically, the phenomenological Chevallier-Polarski-Linder (CPL) parameterization \(w_0w_a\)CDM is often employed\footnote{
It is important to note that while many studies use the simplistic $w_0 w_a$CDM model that allows for dynamical deviations from a cosmological constant \(\Lambda\), it is not necessarily a preferred model, and other evolving dark energy models are often examined (e.g. \citealt{Brownsberger_2019,Dhawan_2020,rebouças2024investigatinglatetimedarkenergy,camilleri2024dark}).
} 
to explore evolving dark energy cosmologies, wherein the dark energy equation of state
\begin{equation}
    w(a) = w_0 + w_a(1 - a) \,,
\end{equation}
where $w_0$ is the value of the equation of state today (or at $a=1$), and $w_a$ describes the rate of change of the equation of state with the scale factor $a$. Both $w_0$ and $w_a$ are constants.
Analyses using \(w_0w_a\)CDM show deviations from the \(\Lambda\)CDM framework at significant levels when combining the DESI BAO \citep{desiy1} with SNe: 2.5\(\sigma\) using Pantheon+ \citep{brout2022pantheon+}, 3.5\(\sigma\) with Union3 \citep{rubin2023union}, and 3.9\(\sigma\) with DES \citep{DESY5}. With DESI 2024 VII full shape results, the combined probes' preference for $w_0>-1$ and $w_a<0$ is reinforced \citep{DESIFS}. These observations align with findings from \cite{berghaus2024quantifying}, who explored scalar-field models of dark energy using DESI BAO data in combination with CMB and SNe measurements. They found that models allowing for a small but non-zero kinetic energy in scalar fields can reduce the tension in the matter density parameter \(\Omega_m\) between BAO and SNe datasets.

Motivated by DESI Y1 results \citep{desiy1}, we aim to investigate the question of whether a true \(w_0w_a\)CDM cosmology could naturally manifest the observed differences in \(\Omega_m\) and $H_0$ when SNe, BAO, and CMB datasets are analyzed separately under the \(\Lambda\)CDM framework, by generating mock $w_0w_a$CDM cosmological data. While it is expected that assuming an incorrect expansion history can lead to biases in the inferred values of cosmological parameters like \( \Omega_m \) and \( H_0 \), the crucial question is whether the specific discrepancies observed in the real datasets are consistent with what would be expected if the true underlying cosmology deviates from \( \Lambda \)CDM and follows the \( w_0w_a \)CDM model preferred by combined data.

Ultimately, we aim to clarify whether the $\Omega_m$ tensions between SNe, BAO, and CMB datasets signify new physics that may be explained by dynamical dark energy. 

This paper is organised as follows. 
In Section \ref{sec:sims}, we introduce our simulations of SNe, BAO, and CMB respectively. 
In Section \ref{sec:backward}, we discuss our methodology and show results. 
Finally in Section \ref{sec:discussion}, we discuss the implications of this work.
Throughout, we assume units km/s/Mpc for $H_0$.

\section{\label{sec:sims}Simulations and Likelihood}

To investigate the impact of cosmological model assumptions on parameter estimation, we generate mock datasets for type Ia supernovae (SNe), baryon acoustic oscillations (BAO), and the cosmic microwave background (CMB). 
For simplicity, we construct idealized mock datasets with realistic uncertainties, but without adding observational noise or scatter. 
This approach allows us to focus on the effects of cosmological models on the data without the complications introduced by statistical fluctuations or selection effects. 
Throughout this paper we assume flatness.
 
\subsection{Type Ia supernovae}\label{subsec:SNe_mocks}

\textbf{Mock Data:} We simulate a dataset of 1,829 SNe Ia, using the same redshifts as the Dark Energy Survey Supernova Program Year 5 (DES-SN5YR) sample \citep{DESY5}. 
This dataset covers a redshift range of $0.02 < z < 1.12$ and provides a DES Y5-like sample as the SNe probe. 

\textbf{Simulation Method:} The simulated SNe Ia data are generated by computing the distance modulus $\mu(z)$ for each supernova using the background cosmological model. 
The distances correspond exactly to the model without added scatter, ensuring that any deviations in parameter estimation arise solely from the model assumptions. 
The distance modulus (or luminosity distance $d_L(z)$) is calculated as 
\begin{equation} 
    \mu(z) = m - M = 5 \log_{10} \left( \frac{d_L(z)}{10 \text{pc}} \right) \,,
\end{equation} 
where for the DES (uncalibrated) SNe, the measured quantity is the apparent magnitude $m$ and where the intrinsic absolute magnitude $M$ of a SN is unknown. 

Thus $M$ is a parameter that is fully degenerate with $H_0$ and thus both parameters ($M$, $H_0$) are marginalized over in the likelihood fit. 

By following this methodology, our analysis closely imitates the DES-SN5YR likelihood, as standard practices in SNe cosmology incorporate these nuisance parameters. 

\textbf{Uncertainties:} We adopt the measurement uncertainties on $\mu(z)$ for each SN and covariance $\mathbf{C_{DES}}$ from the public DES-SN5YR dataset itself \citep{DESY5}.

\textbf{Likelihood:} To account for uncertainties in the absolute magnitude \(M\) and the Hubble constant \(H_0\), we introduce these as additional free parameters and marginalize over \(M\) in the likelihood function. Following the methodology outlined in \cite{DESY5}, the likelihood function for the SNe Ia is defined as:

\begin{equation}
\label{eq:snelikelihood}
\ln \mathcal{L}_{\rm SNe} =
-\frac{1}{2} \left[
\boldsymbol{\Delta \mu}^T \cdot \mathbf{C_{\text{DES}}}^{-1} \cdot \boldsymbol{\Delta \mu}
+ \ln \left( (2\pi)^N |\mathbf{C_{\text{DES}}}| \right)
\right] \,,
\end{equation}
where \( \Delta \mu\) represents the residuals vectors between the model predictions and the observed (mock) data. This modified likelihood function allows us to analytically marginalize over the nuisance parameter \( M \), effectively removing its dependence from the cosmological parameter estimation.

\subsection{Baryon acoustic oscillations\label{BAO_data}}

\textbf{Mock Data:} The BAO data are simulated to match observations from DESI Y1 results \citep{desiy1}. 
This includes measurements from 7 different tracers across different effective redshifts, providing constraints on the expansion rate and geometry of the Universe.

\textbf{Simulation Method:} We compute the BAO observables, specifically the angle-averaged distance $D_V(z)$ and the Alcock-Paczyński parameters, ratio of transverse to line-of-sight comoving distances $F_{\text{AP}}(z)$ from the cosmological model.
The angle-averaged distance is 
\begin{equation} 
    D_V(z) = \left[ z D_M^2(z) D_H(z) \right]^{1/3} \,,
\end{equation} 
where $D_M(z)$ is the comoving angular diameter distance, and $D_H(z) = c/H(z)$ is the Hubble distance. The comoving distances and Hubble parameters are calculated using \texttt{astropy.cosmology} \citep{astropy}.
The ratio $F_{\text{AP}}(z)$ of transverse to line-of-sight comoving distances  is
\begin{equation} 
    F_{\text{AP}}(z) = \frac{D_M(z)}{D_H(z)} \,.
\end{equation}

We note that our method uses $D_V(z)/r_d$ and $F_{\text{AP}}(z)$, while DESI Y1 uses $D_M/r_d$, $D_H/r_d$ and $D_V/r_d$, but they are equivalent. 
For the drag-epoch sound horizon $r_d$, we use \footnote{
We have also incorporated a direct calibration of \( r_d \); please see Appendix~\ref{sec:appendix_rd}.
} Eq. (2.5) from the DESI Y1 cosmology paper \cite{desiy1}. 
We apply a Gaussian distribution from BBN on $\Omega_bh^2=0.02218\pm0.00055$ \citep{BBN} in the MCMC analysis in Fig.~\ref{fig:bestw0wa_to_lcdm}. 
However, for the Hessian matrix maximization analyses in this work (used for Fig.~\ref{fig:w0wa_to_lcdm_all}), for computational feasibility, we use the central value $\Omega_bh^2=0.02218$ from BBN constraints \citep{BBN}. 
See each method in Section~\ref{sec:backward_method} for details.

\textbf{Uncertainties:} Uncertainties and correlations between $D_V$, $D_M$, and $D_H$ are adopted from Table 1 of the DESI Y1 cosmology paper \citep{desiy1}. 
For tracers where both $D_M$ and $D_H$ are provided, we calculate the uncertainties in $D_V$ and $F_{\text{AP}}$ using error propagation formulas, accounting for the correlation coefficient $r$ between $D_M$ and $D_H$.
The uncertainty in $D_V$ is
\begin{equation} 
    \sigma_{D_V} = \frac{D_V}{3} \sqrt{ 4 \left( \frac{\sigma_{D_M}}{D_M} \right)^2 + \left( \frac{\sigma_{D_H}}{D_H} \right)^2 + 4 r \frac{\sigma_{D_M}}{D_M} \frac{\sigma_{D_H}}{D_H} } \,.
\end{equation}
Similarly, the uncertainty in $F_{\text{AP}}$ is 
\begin{equation} 
    \sigma_{F_{\text{AP}}} = F_{\text{AP}} \sqrt{ \left( \frac{\sigma_{D_M}}{D_M} \right)^2 + \left( \frac{\sigma_{D_H}}{D_H} \right)^2 - 2 r \frac{\sigma_{D_M}}{D_M} \frac{\sigma_{D_H}}{D_H} } \,,
\end{equation}
where $r$ represents the correlation between \(D_M\) and \(D_H\) given in Table 1 of \cite{desiy1}.

For the 5 of 7 tracers from DESI where both \(D_M\) and \(D_H\) are available, we use both \(D_V(z)/r_d\) and \(F_{AP}(z)\). 
For the remaining 2 tracers with only \(D_V\) provided, we use just \(D_V(z)/r_d\) as basis of the data for simulations.

\textbf{Likelihood:} The likelihood function for BAO is defined as:
\begin{equation}
\label{eq:baolikelihood}
    \ln \mathcal{L}_{\rm BAO} = -\frac{1}{2} \left[ \sum_i \left( \frac{M_i - T_i}{\sigma_i} \right)^2 + \sum_i \ln (2\pi \sigma_i^2) \right] \,,
\end{equation}
with observed (or mock) BAO data \( M_i \) for \(D_V/r_d\) and \(F_{AP}\), theoretical predictions \( T_i \) from the cosmological model, and uncertainties \(\sigma_i\) from DESI Y1 papers \citep{desiy1}.

\subsection{Cosmic Microwave Background}\label{subsec:cmb_mocks}

\textbf{Mock Data:} We incorporate constraints from CMB using two parameters - the acoustic scale $l_A$ and the shift parameter $R$. The uncertainties on these parameters and the physical baryon density used in our calculations are derived from the Planck 2018 results and covariance \citep{Planck2018}. 

\textbf{Simulation Method:} The shift parameter $R$ and the acoustic scale $l_A$ are calculated as \cite{komatsu2009five}
\[
R = \sqrt{\Omega_m H_0^2} \frac{D_M(z_*)}{c} \label{eq:R_shift} \,,
\]
\[
l_A = (1 + z_*) \frac{\pi D_M(z_*)}{r_s(z_*)} \label{eq:lA_acoustic} \,,
\]
where $D_M(z_*)$ is the comoving angular diameter distance to the photon-decoupling redshift $z_*$, $r_s(z_*)$ is the comoving sound horizon at $z_*$, and $c$ is the speed of light. 
Note that $r_{z_*}$ is physically different from the drag-epoch sound horizon $r_d$ in the BAO section.

\begin{figure}
    \centering
    \includegraphics[width=\linewidth]{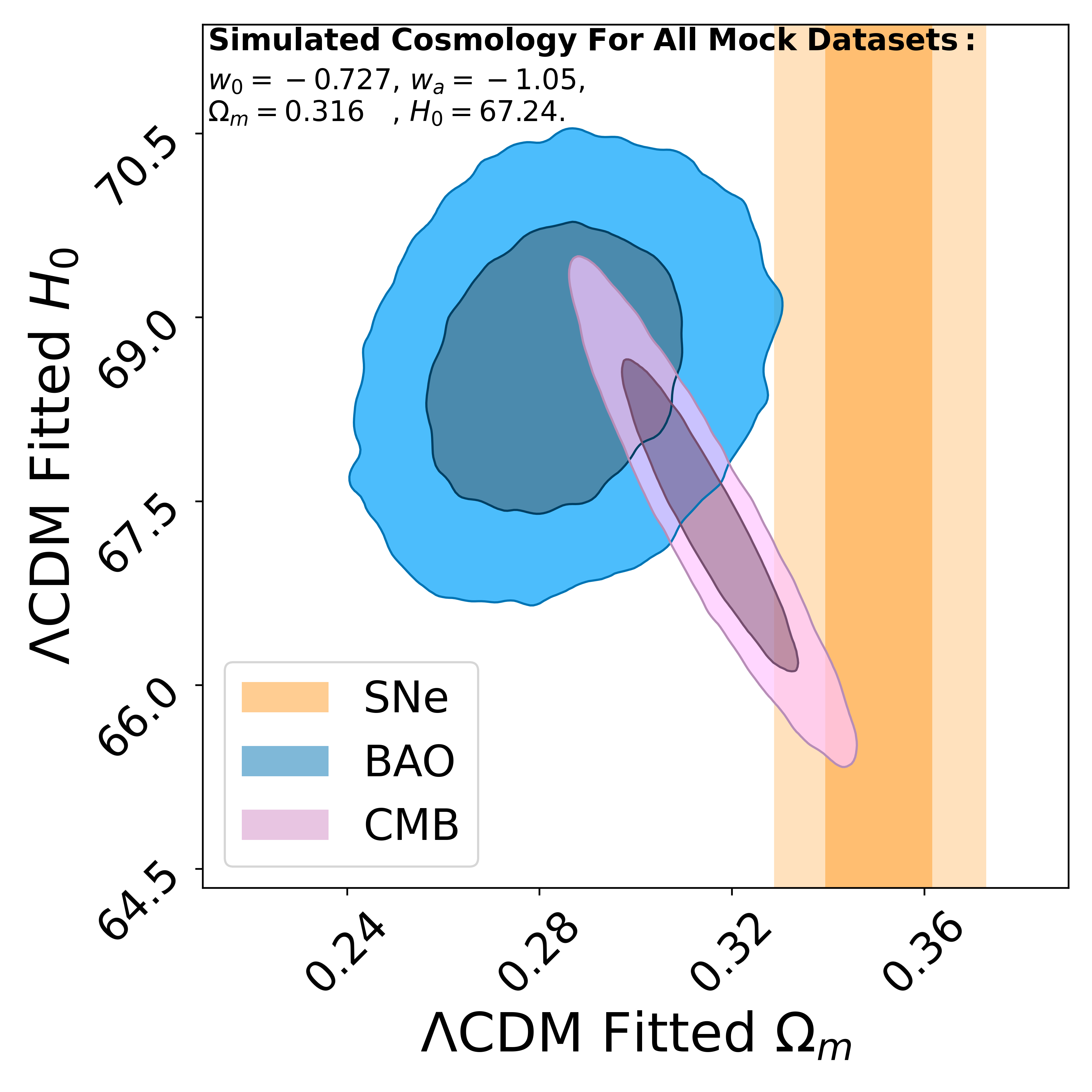}
    \caption{
    Constraints on \(\Omega_m\) and \(H_0\) obtained by fitting \(\Lambda\)CDM to mock CMB, BAO, and SNe datasets generated in the DESI+DES-SN+Planck \citep{desiy1} best-fit \(w_0w_a\)CDM cosmology \( \{w_0=-0.727, w_a=-1.05, \Omega_m=0.316, H_0=67.24\} \). 
    The contours represent the 68\% and 95\% confidence regions for each mock dataset individually: SNe (yellow), BAO (blue), and CMB (pink). 
    This figure illustrates the \(\Omega_m\) and \(H_0\) discrepancies between different probes when fitting \(\Lambda\)CDM to data simulated in a dynamical \(w_0w_a\)CDM cosmology.}
    \label{fig:bestw0wa_to_lcdm}
\end{figure}

The redshift of photon decoupling, $z_*$, is calculated using the fitting formula from Eq. (8) from \cite{chen2019distance}, which takes the physical baryon density $\Omega_b h^2$ and \(\Omega_m\) as input and can be adjusted to the chosen cosmology. 
As done for BAO mocks (Section~\ref{BAO_data}), for computational feasibility, we take the BBN constraint of $\Omega_b h^2 = 0.02218$ \citep{BBN} for our Hessian matrix maximization analyses for Fig.~\ref{fig:w0wa_to_lcdm_all}, whereas for the MCMC analysis done in Fig.~\ref{fig:bestw0wa_to_lcdm}, we use a Gaussian BBN prior on $\Omega_bh^2=0.02218\pm0.00055$ \citep{BBN}.

The comoving angular diameter distance and the comoving sound horizon $r_s(z_*)$ at $z_*$ is computed using numerical integration routines from \texttt{SciPy} \citep{scipy2020} and \texttt{Astropy} \citep{astropy} with Equation 3 from from \cite{chen2019distance}.

\textbf{Uncertainties:} 
We adopt the uncertainties and covariance between $R$ and $l_A$ from the Planck 2018 results \citep{Planck2018}. The covariance matrix is given by:
\begin{equation}
    \mathbf{C_{P18}} = \begin{pmatrix}
    \sigma_R^2 & \sigma_{Rl_A} \\
    \sigma_{Rl_A} & \sigma_{l_A}^2
\end{pmatrix} \,, \label{eq:covariance}
\end{equation}
with the variances and covariance specified from the Planck 2018 TT,TE,EE+lowE likelihoods shown in Table \ref{tab:planck_results}.

\begin{table}[t]
\centering
\begin{minipage}[t]{0.48\textwidth}
\centering
\begin{tabular}{lc}
\hline
 & Parameter Uncertainty \\
 & $\sigma$ \\
\hline
$R$   & $0.0065$ \\
$l_A$ & $0.105$ \\
\hline
\end{tabular}
\end{minipage}
\hfill
\begin{minipage}[t]{0.48\textwidth}
\centering
\begin{tabular}{lcc}
\hline
 & \multicolumn{2}{c}{Correlation Coefficients} \\
 & $R$ & $l_A$ \\
\hline
$R$   & 1.0   & $0.47$ \\
$l_A$ & $0.47$ & 1.0   \\
\hline
\end{tabular}
\end{minipage}
\caption{Uncertainties (standard deviations) (top) and correlation coefficients (bottom) between the shift parameter $R$ and the acoustic scale $l_A$ from the Planck 2018 TT, TE, EE + lowE analysis for the $\Lambda$CDM cosmological model \citep{Planck2018}.}
\label{tab:planck_results}
\end{table}

Since we are generating idealized simulated data, the observed values $R_\text{obs}$ and $l_{A,\text{obs}}$ correspond exactly to the theoretical values computed from the underlying cosmological model without added noise.

\textbf{Likelihood:} The likelihood function for the CMB is computed by comparing the theoretical values of $R$ and $l_A$ with the observed values, incorporating the covariance between these parameters. The negative log-likelihood is given by
\begin{equation}
    \ln \mathcal{L}_{\rm CMB} = -\frac{1}{2} \left[ \Delta \mathbf{d}^\mathrm{T} \cdot \mathbf{C_{P18}}^{-1} \cdot \Delta \mathbf{d} + \ln \left( 2\pi |\mathbf{C_{P18}}| \right) \right] \,,\label{eq:cmblikelihood}
\end{equation}
where $\Delta \mathbf{d}$ is the difference vector
\begin{equation}
    \Delta \mathbf{d} = \begin{pmatrix}
    R_\text{theory} - R_\text{obs} \\
    l_{A,\text{theory}} - l_{A,\text{obs}}
    \end{pmatrix} \,.
\end{equation}

\section{\textbf{Analysis}\label{sec:backward}}

\begin{figure}
    \centering
    \includegraphics[width=\linewidth]{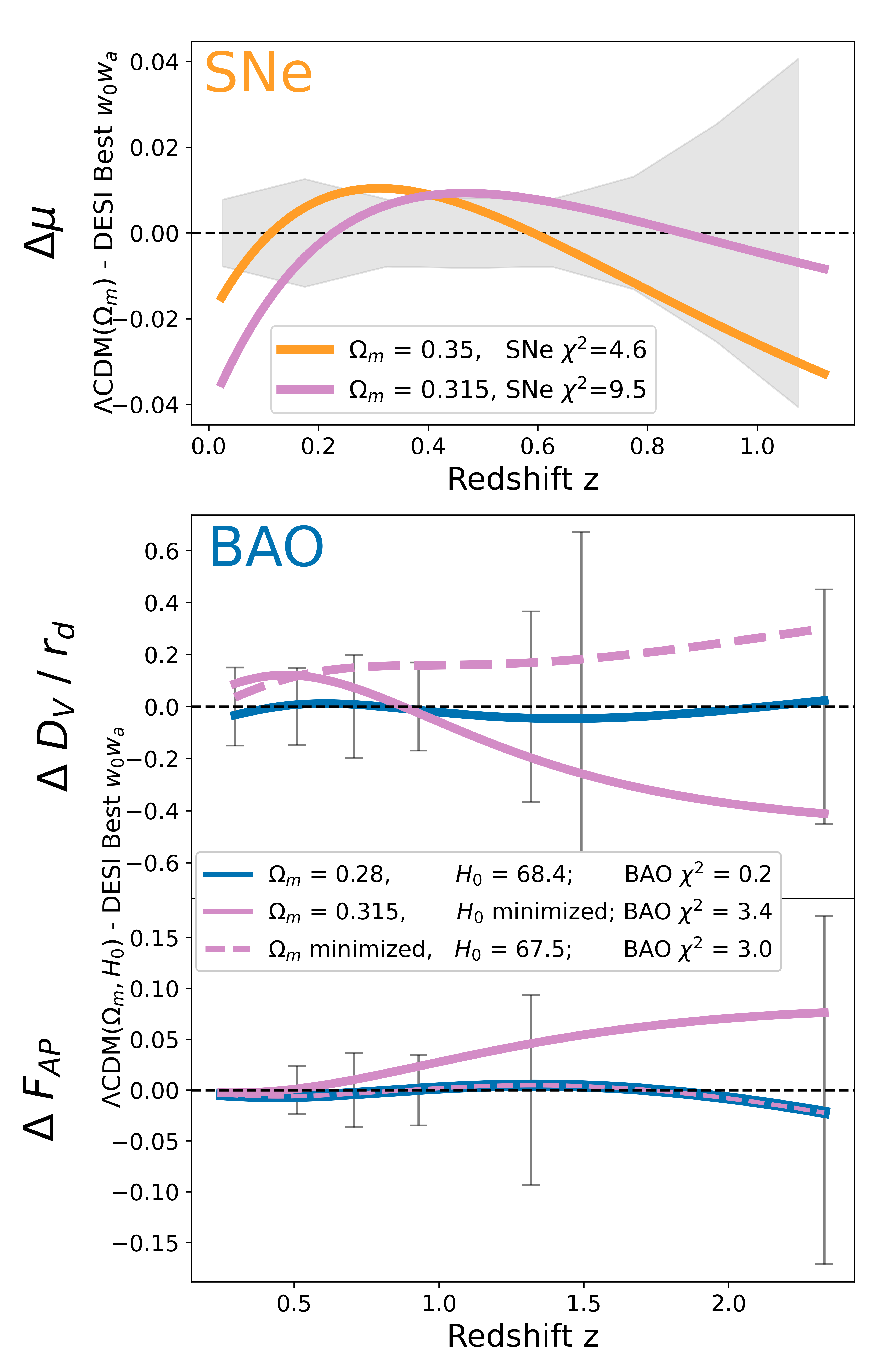}
    \caption{
    Comparison between the best-fit $\Lambda$CDM cosmology (with $\Omega_m = 0.35$ for SNe or $\Omega_m = 0.28$ for BAO) and a Planck-like $\Lambda$CDM cosmology ($\Omega_m = 0.315$) against the DESI+CMB+DESY5SN \citep{desiy1} best-fit $w_0w_a$CDM model $\{w_0 = -0.727$, $w_a = -1.05$, $\Omega_m = 0.316$, $H_0 = 67.24\}$ with realisitc data uncertainties shown for SNe (top) and BAO (bottom). The three panels show residuals in the SNe distance modulus $\mu$, the BAO $D_V / r_d$, and the BAO $F_{AP}$ versus redshift $z$. Error bars (shaded regions) reflect uncertainties in DES-SN5YR \citep{DESY5} and DESIY1 \citep{desiy1}, with binned SNe errors using weighted averages of uncertainties in redshift bins of width $\Delta z=0.15$ for illustration. The $\chi^2$ values reported indicate goodness-of-fit for each model (colored curves) relative to the mock data generated in $w_0w_a$CDM (horizontal dashed), calculated similarly to Equation~\ref{eq:snelikelihood} and \ref{eq:baolikelihood} (without normalization term). 
    `Minimized' refers to floating one parameter while fixing others to achieve the lowest $\chi^2$. For BAO, when $\Omega_m = 0.315$, $H_0$ is `minimized' to 68.8; when $H_0 = 67.5$, $\Omega_m$ is `minimized' to 0.279. 
    For the SNe, $H_0$ is also `minimized' for both  models presented.}
\label{fig:mu_fap_dv_lcdm_w0wabest}
\end{figure}

In this section, we explore how discrepancies in \(\Omega_m\) and \(H_0\) between different cosmological probes might manifest if the true underlying cosmology is \(w_0w_a\)CDM but it is analyzed within the \(\Lambda\)CDM framework. 
Our analysis consists of two main parts. 
First, we examine the single dynamical dark energy cosmology derived in \cite{desiy1} from the best-fit results of the combined DESI+DES+Planck data. 

Second, to evaluate the probabilities of occurrences of differences in $\Omega_m$ and $H_0$ across the different probes, we sample from the entire observed posterior distribution of the \(w_0w_a\)CDM model. We then compare these differences among the different probes to the discrepancy found in the real datasets.

\subsection{\label{sec:backward_method}Single Realization from a Best-Fit \(w_0w_a\)CDM}

In this section, we first calculate the probability of obtaining the observed constraints in \(\Lambda\)CDM for each probe individually if the true underlying cosmology is \(w_0w_a\)CDM. 
First, we create mock datasets \{SNe, BAO, CMB\} in the DESI+DES+Planck \(w_0w_a\)CDM best fit cosmology \{$w_0=-0.727$, $w_a=-1.05$, $\Omega_m=0.316$, $H_0=67.24$\}. We then fit each mock dataset individually in \(\Lambda\)CDM with an MCMC, applying a Gaussian distribution on the physical baryon density \(\Omega_b h^2 = 0.02218 \pm 0.00055\) from BBN \citep{BBN} for the BAO and CMB fits. 
We float \(\Omega_m\) and \(H_0\) as free parameters, while \(\Omega_b h^2\) is constrained by the prior. 
As shown in Fig.~\ref{fig:bestw0wa_to_lcdm} and Table~\ref{tab:background_vs_fitted}, we find that the best-fit values of \(\Omega_m\) under \(\Lambda\)CDM for each dataset are $\Omega_{m, \text{SNe}} = 0.35 \pm 0.012, \quad \Omega_{m, \text{BAO}} = 0.28 \pm 0.018, \quad \Omega_{m, \text{CMB}} = 0.315 \pm 0.012$. These values are consistent with those observed in the real datasets, as reported by DES-SN5YR SNe (\(\Omega_m = 0.353 \pm 0.017\)) \cite{DESY5}, DESI BAO (\(\Omega_m = 0.295 \pm 0.015\)) \cite{desiy1}, and Planck 2018 CMB (\(\Omega_m = 0.3166 \pm 0.0084\)) \cite{Planck2018}. 

In order to illustrate why dynamical dark energy results in higher $\Omega_m$ for SNe and lower $\Omega_m$ for BAO, we show data vector residuals relative to the DESI \citep{desiy1} maximum likelihood \(w_0w_a\)CDM cosmology in Fig.~\ref{fig:mu_fap_dv_lcdm_w0wabest}. 
For each probe (top: SNe, bottom: BAO), we show data vectors corresponding to both the best-fit \(\Lambda\)CDM cosmological parameters and also corresponding to Planck-like \(\Lambda\)CDM cosmological parameters (i.e. $\Omega_m=0.315$).
SN data favor a higher \(\Omega_m\) to match the observed (uncalibrated) luminosity distances, while BAO data prefer a lower \(\Omega_m\) to fit the observed $D_V/r_d$ and $F_{AP}$. Fig.~\ref{fig:mu_fap_dv_lcdm_w0wabest} also illustrates the effect of dynamical dark energy leading to the BAO preference (lower $\chi^2$) for higher $H_0$ values in $\Lambda$CDM, as the curve with a slightly higher $H_0 = 68.4$ has the smallest $\chi^2 = 0.2$. 

\subsection{Realizations from Posterior Samples of \(w_0w_a\)CDM \label{sec:backward_method_many}}
In this section we compare the amount of disagreement between real data probes analyzed individually in \(\Lambda\)CDM with the expected disagreement from mocks simulated in all viable \(w_0w_a\)CDM cosmologies to understand if the real observed disagreement can be entirely explained by \(w_0w_a\)CDM. 
To ensure that the constraints on \(\Omega_m\) shown in Fig.~\ref{fig:bestw0wa_to_lcdm} and the second row of Table~\ref{tab:background_vs_fitted} are not statistical flukes and to evaluate the probability of observing each \(\Omega_m\) individually, we utilize the posterior DES24 chains provided\footnote{
At the time of this work, the DESI chains are not yet public. 
} 
in \cite{DESY5} for the \(w_0w_a\)CDM model. 
Specifically, we use the combined probe constraints (DES24) in \(w_0w_a\)CDM. 
For each set of cosmological parameters \{\(w_0, w_a, \Omega_m, H_0\)\} from the DES24 chains, we generate mock SNe, BAO and CMB datasets with the methods described in Section \ref{sec:sims}, setting \(\Omega_bh^2=0.02218\), the centeral value from BBN \cite{BBN}, for CMB and BAO mock datasets. 
All three probes are simulated in the same \(w_0w_a\) cosmology. 
Each probe is then analyzed individually within the \(\Lambda\)CDM framework, treating both \(H_0\) and \(\Omega_m\) as free parameters.

The likelihood is maximized to find the best-fit values of \(\Omega_m\) and \(H_0\) for each of the SNe, BAO and CMB datasets generated in \(w_0w_a\)CDM. 
Although our mock SNe datasets are not sensitive to \(H_0\), we nonetheless float it simultaneously with \(\Omega_m\) to match the methodology of \cite{DESY5}. 
Because this maximum likelihood is calculated thousands of times for each of the cosmologies in the DES24 chains, we simply calculate the maximum likelihood point and then compute the Hessian matrix ($\mathbf{H}$) at the best-fit values to estimate the uncertainties in \(\Omega_m\) and \(H_0\) as 
\begin{equation}
    \mathbf{H} = 
    \begin{pmatrix}
        \frac{\partial^2 (-\ln \mathcal{L})}{\partial \Omega_m^2} & \frac{\partial^2 (-\ln \mathcal{L})}{\partial \Omega_m \partial H_0} \\
        \frac{\partial^2 (-\ln \mathcal{L})}{\partial H_0 \partial \Omega_m} & \frac{\partial^2 (-\ln \mathcal{L})}{\partial H_0^2}
    \end{pmatrix} \,, \nonumber
\end{equation}
\begin{equation}
    \quad
    \begin{matrix}
        \sigma_{\Omega_m} &= \sqrt{ (\mathbf{H}^{-1})_{11} } \\
        \sigma_{H_0} &= \sqrt{ (\mathbf{H}^{-1})_{22} }
    \end{matrix} \,.
\end{equation}

\begin{figure*}
    \centering
    \includegraphics[width=\textwidth]{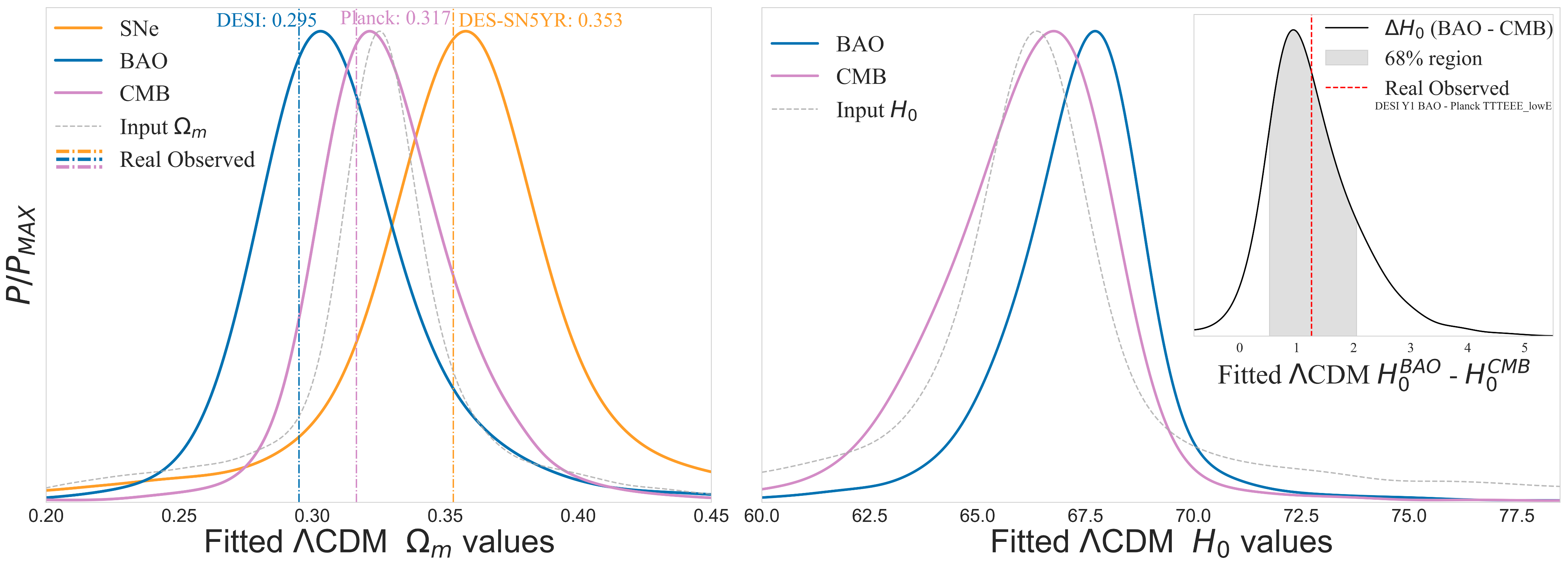}
    \caption{The figure shows marginalized \(\Lambda\)CDM fits for \(H_0\) and \(\Omega_m\) to mock datasets in \(w_0w_a\)CDM cosmologies. Simulated \(w_0w_a\)CDM sample the public DES24 chains. 
    \textbf{Left panel:} The $\Lambda$CDM \(\Omega_m\) distributions for various probes \{SNe, BAO, and CMB\}. 
    The gray dashed curve shows the underlying values of $\Omega_m$ used for these simulations. 
    The three vertical lines represent the observed \(\Lambda\)CDM \(\Omega_m\) values for real data for reference: DESI-Y1 \citep{desiy1} (blue), DES-SN5YR \citep{DESY5} (yellow), and Planck18 CMB only \citep{Planck2018} (pink). 
    \textbf{Right panel:} The $\Lambda$CDM \(H_0\) distributions for BAO and CMB probes. 
    The gray dashed curve shows the \(w_0w_a\)CDM \(H_0\) values used to simulate the mocks. 
    SNe are not shown here because DES-like SNe are not sensitive to \(H_0\) by themselves, but require an external calibration. 
    \textbf{Right sub-panel:} The corresponding \(\Delta H_0\) distribution, quantifying the difference between \(H_0\) results from BAO and CMB datasets at each step in the chains. 
    The shaded region represents the 68\% credible interval. 
    The red dashed line indicates the observed \(\Delta H_0\) between DESI-Y1 BAO and Planck18 CMB under \(\Lambda\)CDM. 
    Our observed difference between the simulated \(w_0w_a\)CDM maximum likelihood \(H_0\) and the `observed' mock-BAO \(\Lambda\)CDM maximum likelihood \(H_0\) is 1.2 km/s/Mpc.}
    \label{fig:w0wa_to_lcdm_all} \label{fig:H0_cmb_bao}
\end{figure*}

Finally, to build a probability distribution for the \(\Lambda\)CDM fits, we assume each of the free parameters follows a Gaussian distribution $P^i(\Omega_{m, \text{probe}}^{\Lambda \text{CDM},i})$ for every background \(w_0w_a\)CDM cosmology $i$. 
These probabilities are then summed to produce the posterior density of \(\Lambda\)CDM parameters $P(\Omega_{m, \text{probe}}^{\Lambda \text{CDM}})$, given all of the background cosmologies from the DES24 chains. 
This process is mirrored to build the posterior probabilities $P(H_{0,\text{probe}}^{\Lambda \text{CDM}})$ for the Hubble Constant $H_0$. 

To quantify the statistical significance of the discrepancies in \(\Omega_m\) among the probes, we perform a $\chi^2$ test for consistency. 
For each set of \(\Omega_m\) values obtained from the \(\Lambda\)CDM fits to the mock datasets, 
we calculate a $\chi^2$ statistic for the fit under the assumption that all of the individual probes' measurements $i$ are a random statistical fluctuation of a single $\Omega_{m,\text{True}} = \frac{\sum_{i} \Omega_{m,i} / \sigma_{i}^2}{\sum_{i} 1 / \sigma_{i}^2}$. 
We then quantify the probability that all probes measure the same true $\Omega_{m,\text{True}}$, i.e. the probability of obtaining a $\chi^2$ value as extreme or more extreme than the observed \(\chi^2\dagger\) under the null hypothesis that all probes measure the same true $\Omega_{m,\text{True}}$. 
This is given by  $\chi^{2}\dagger = \sum_{i} \frac{ (\Omega_{m,i} - \Omega_{m,\text{True}} )^2 }{ \sigma_{i}^2 }$. 
We then compute the \(p\text{-value}\) using the $\chi^2$ cumulative distribution function with $\nu=2$ degrees of freedom 
\begin{equation}
    \label{eq:p value}
    p\text{-value} = 1 - \text{CDF}_{\chi^2}(\chi^2\dagger, \nu) \,.
\end{equation}
A smaller \(p\text{-value}\) indicates that the discrepancies among the probes are less likely to occur by chance.
% if they share the same \(\Omega_m\).

Using the best-fit \(\Omega_m\) values and uncertainties reported by DES-SN5YR \(\Omega_m = 0.353 \pm 0.017\) \citep{DESY5}, DESI-Y1 BAO \(\Omega_m = 0.295 \pm 0.015\) \citep{desiy1}, and Planck 2018 CMB \(\Omega_m = 0.315 \pm 0.007\) \citep{Planck2018}, the observed \(p\text{-value}\) is 0.035, shown in Table~\ref{tab:background_vs_fitted}. In a frequentist sense, this means that if these three measurements with given uncertainties were all drawn from a single true value of \(\Omega_m\) under \(\Lambda\)CDM, only about 3.5\% of random realizations would exhibit discrepancies as large or larger than what is observed. Also note, $p\text{-value} = 0.013$ if we calculate using the DESI Y1 full-shape analysis \citep{DESIFS} \(\Omega_m = 0.2962 \pm 0.0095\) instead of the DESI Y1 BAO \citep{desiy1}.

\subsection{\label{sec:backward_results}Results}

\begin{table*}[t]
    \centering
    \resizebox{\textwidth}{!}{%
    \begin{tabular}{?c|c|c?}
        \thickhline
        \multicolumn{1}{?c|}{\begin{tabular}[c]{@{}c@{}}Data/Mock\end{tabular}} &
        \multicolumn{1}{c|}{\begin{tabular}[c]{@{}c@{}}$\Lambda$CDM Fit\end{tabular}} &
        \multicolumn{1}{c?}{\begin{tabular}[c]{@{}c@{}}\phantom{A}$\Omega_m$ Agreement Between Probes\phantom{A} \\ (See Section III.1)\end{tabular}} \\ \thickhline 
        \multicolumn{1}{?c|}{\begin{tabular}[c]{@{}c@{}}Real Data \\ (DESI Y1 VI BAO, DES-SN5YR, Planck18 CMB)\end{tabular}} &
        \begin{tabular}[c]{@{}l@{}}\vspace{.01in} \\BAO: $\Omega_m=0.295{\pm0.015}$, $H_0=68.5\pm{0.8}$ \\ 
        SNe : $\Omega_m=0.353{\pm0.017}$ \\  
        CMB: $\Omega_m=0.315{\pm0.007}$, $H_0=67.3\pm{0.6}$ \\
        \vspace{.01in} \end{tabular} &
        $p$-value = 0.035 \\ \thickhline 
        \multicolumn{1}{?c|}{\begin{tabular}[c]{@{}c@{}}Mock simulated in \\ DESI+CMB Best-Fit $\Lambda$CDM \\ $\Omega_m=0.31$, $H_0=68$\end{tabular}} &
        \begin{tabular}[c]{@{}l@{}}\vspace{.03in} \\BAO: $\Omega_m=0.311\pm0.019$, $H_0=68.0 \pm 0.8$ \\ 
        SNe : $\Omega_m=0.310 \pm 0.011$ \\ 
        CMB: $\Omega_m=0.310\pm0.012$, $H_0=68.0\pm 0.8$ \\ \vspace{.04in}\end{tabular} &
        $p$-value = 0.999 \\  \thickhline 
        \multicolumn{1}{?c|}{\begin{tabular}[c]{@{}c@{}}Mock simulated in \\ DESI+CMB+DESY5SN Best-Fit $w_0w_a$CDM \\ $\Omega_m=0.316$, $w_0=-0.727$, $w_a=-1.05$, $H_0=67.24$\end{tabular}} &
        \begin{tabular}[c]{@{}l@{}}\vspace{.03in} \\BAO: $\Omega_m=0.281^{+0.019}_{-0.016}$, $H_0=68.6\pm0.8$ \\ 
        SNe : $\Omega_m=0.350\pm 0.011$ \\ 
        CMB: $\Omega_m=0.315\pm 0.012$, $H_0=67.4\pm0.8$ \\\vspace{.04in}\end{tabular} &
        $p$-value = 0.003 \\  \thickhline 
        \multicolumn{1}{?c|}{\begin{tabular}[c]{@{}c@{}} Mocks simulated for each step in DES24\\ (DES-SNY5 + eBOSS + Planck18 + 3x2pt) \\ $w_0w_a$CDM chain
        \end{tabular}} &
        \begin{tabular}[c]{@{}l@{}} 
        \vspace{.03in} \\BAO: $\Omega_m=0.303^{+0.051}_{-0.043}$, $H_0=67.8^{+1.3}_{-1.8}$ \\ 
        SNe : $\Omega_m=0.355^{+0.055}_{-0.049}$ \\
        CMB: $\Omega_m=0.321^{+0.030}_{-0.023}$, $H_0=66.9^{+1.5}_{-2.3}$ \\\vspace{.04in}\end{tabular} &
        68\% of $ p $-values within [$9\times10^{-6},0.17$]\\ \thickhline
    \end{tabular}
    }
    \caption{Comparison of the \(\Lambda\)CDM fits to real and mock data, and the \( p \)-values assessing the agreement of \(\Omega_m\) among different probes, as described in Eq.~\eqref{eq:p value}. A smaller \(p\text{-value}\) indicates more discrepancies on \(\Omega_m\) among probes. 
    Each row corresponds to a different data/mock scenario. \textbf{Row 1:} Results from fits to real observational data \citep{desiy1, DESY5, Planck2018}, providing a baseline for comparison with mock scenarios.  
    \textbf{Row 2:} A null test with mocks simulated under a common \(\Lambda\)CDM cosmology, showing a \( p \)-value close to 1, indicating agreement between fitted $\Omega_m$. 
    \textbf{Row 3:} Same as described in Section~\ref{sec:backward_method} and illustratedin Fig.~\ref{fig:bestw0wa_to_lcdm} and \ref{fig:mu_fap_dv_lcdm_w0wabest}. \textbf{Row 4:} The 68\% credible interval of best-fit parameters for fits to the DES24 chain cosmologies, which is also described in Section~\ref{sec:backward_method_many} and illustrated in Fig.~\ref{fig:w0wa_to_lcdm_all}.}
    \label{tab:background_vs_fitted}
\end{table*}

\centerline{ \textbf{Matter density $\Omega_m$} } 

Our analysis reveals that the best-fit \(\Omega_m\) values obtained from individual \(\Lambda\)CDM fits to the mock datasets differ across the probes when the true underlying cosmology is a dynamical dark energy. 
These \(\Omega_m\) values are also observed in real data \citep{desiy1}. 
As shown in Table~\ref{tab:background_vs_fitted} row 4 (DES-SN5YR+CMB+BAO+3x2pt \(w_0w_a\) chains), the mock BAO data prefer a lower \(\Omega_m\), the mock CMB data yield an intermediate \(\Omega_m\), and the mock SN data favor a higher \(\Omega_m\). 
The observed values from DESI-Y1 BAO \citep{desiy1}, Planck 2018 CMB \citep{Planck2018}, and DES-SN5YR \citep{DESY5} are also shown in Fig.~\ref{fig:w0wa_to_lcdm_all} with vertical dash-dotted lines for reference (also in Table~\ref{tab:background_vs_fitted}, row 1). 
Our simulated \(\Omega_m\) distributions closely match the observed values: 
the BAO \(\Omega_m\) distribution peaks around 0.313, similar to the real DESI observed value of 0.295; 
the CMB \(\Omega_m\) distribution centers around 0.323, matching the real Planck value of 0.317; 
and the SNe \(\Omega_m\) distribution peaks near 0.355, consistent with the real DES-SN5YR value of 0.353. 

To quantify the statistical significance of the discrepancies in \(\Lambda\)CDM \(\Omega_m\) among the probes, we performed a \(p\text{-value}\) test for consistency as described in Section~\ref{sec:backward_method} with Eq.~\eqref{eq:p value}, summarized in Table~\ref{tab:background_vs_fitted}. 
For real data, the observed agreement between \(\Omega_m\) has a $p$-value of 0.035 using DESI 2024 VI \cite{desiy1}. With the release of DESI full-shape results, this discrepancy persists \citep{DESIFS}, the $p$-value dropping to 0.013, showing an even larger discrepancy. 

For mock data simulated in the \citep{desiy1} best-fit \(w_0w_a\)CDM cosmology, we find a $p$-value of 0.003. When examining the distributions of \(\Lambda\)CDM \(\Omega_m\) agreement $p$-values for all of the \(w_0w_a\)CDM cosmologies in the DES24 chains, we find that approximately 50.3\% of the simulated \(w_0w_a\)CDM cosmologies have \(p\text{-value}\)s smaller than that of the real observed data $p$-value $<$ 0.035, and that the 68\% $p$-value confidence interval is [$9\times10^{-6},0.17]$. Only $3.1\%$ of the simulated cosmologies have $p > 0.5$. This suggests that if the true underlying \(w_0w_a\)CDM cosmology is indeed well characterized by DES24 \(w_0w_a\)CDM posteriors, the observed discrepancy in \(\Omega_m\) in \(\Lambda\)CDM for the real datasets is entirely likely. 

We also show a null test in the second row of Table~\ref{tab:background_vs_fitted}, where \(\Lambda\)CDM is both the underlying mock cosmology and the assumed cosmology when fitting parameters, and we find agreement between the $\Omega_m$ values and corresponding $p$-value of 0.999.

\centerline{ \textbf{Hubble constant $H_0$} } 

Additionally, because we also simultaneously simulate $H_0$ in the mock datasets following each $H_0$ in the DES24 \(w_0w_a\)CDM chains, we can compare the fitted \( H_0 \) for BAO and CMB in \(\Lambda\)CDM. 
The best-fit values of \( H_0 \) derived from the BAO data are biased higher than the input and the CMB, as shown in Fig.~\ref{fig:H0_cmb_bao}. 
The last row of Table~\ref{tab:background_vs_fitted} shows that our \(\Lambda\)CDM BAO results peak at \( H_{0,\text{BAO}}^{\Lambda \text{CDM}} = 67.7\ \text{km/s/Mpc}\), whereas for CMB, we recover the input \(H_{0,\text{CMB}}^{\Lambda \text{CDM}} = 66.7 \ \text{km/s/Mpc}\). 
The background simulated $H_0$ from the DES24 \(w_0w_a\)CDM chains have a mean of 66.4 $\text{km/s/Mpc}$ (shown by the gray dashed vertical line in Fig.~\ref{fig:H0_cmb_bao}). 
This trend of $\Delta H_0 \sim 1.2$km/s/Mpc is consistent with the findings reported in the DESI collaboration's Year 1 results \citep{desiy1} (also shown in the first row of Table~\ref{tab:background_vs_fitted}), where BAO measurements indicate a higher value of the Hubble constant ($\sim 68.5 \ \text{km/s/Mpc}$) than those derived from \cite{Planck2018} ($\sim 67 \ \text{km/s/Mpc}$).

\section{Discussion \label{sec:discussion}}

This paper set out to investigate whether an underlying dynamical dark energy cosmology could naturally produce the observed differences in \(\Omega_m\) when data are analyzed under the \(\Lambda\)CDM framework. 
Using posterior samples from the chains of the DES Y5 cosmology analysis \citep{DESY5} (to be updated to DESI Y1 when made publicly available), we generated mock datasets based on \(w_0w_a\)CDM cosmologies and then individually fit the various probes with a \(\Lambda\)CDM model. 
We found that the differences in the best-fit \(\Omega_m\) values between SNe, BAO, and CMB datasets closely matches the discrepancies observed in real data. 

These results suggest that the observed discrepancies in \(\Omega_m\) in $\Lambda$CDM between different probes can indeed be explained by an evolving dark energy model. Though, we show in Appendix~\ref{sec:forward} that we cannot rule out the inverse possibility that unknown systematic errors which mimic differences in $\Omega_m$ in $\Lambda$CDM, could be the source of the dynamical dark energy signal. It is thus extremely important to understand if the differences between the probes on \(\Omega_m\) could be systematically driven. 

The current state-of-the-art analyses of SNe (Pantheon+ \citealt{brout2022pantheon+}, DES-SN5YR \citealt{DESY5}, and Union3 \citealt{rubin2023union}) are consistent with high \(\Omega_m\). It is important to note that while the DES and Pantheon+ samples are largely independent, they do share many of the low-$z$ SNe. And while the Pantheon+ and Union3 samples utilize vastly different cosmological pipelines (one frequentist validated \citealt{Armstrong_2023}, another hierarchical Bayesian \citealt{Rubin_2015}), they share many of the same SNe and the underlying SALT model \citep{SALT3}. 
Therefore, while there is agreement between the 3 samples towards higher \(\Omega_m\), there is the possibility of correlated or common systematics. 
Conversely, \cite{brout2022pantheon+} Appendix  Fig. 15 highlights the major updates that have occurred in the last decade of SNe cosmology that have contributed to higher \(\Omega_m\). 
There are roughly equal contributions from 1) updating the Milky Way dust color law from CCM \citep{Cardelli89} to Fitzpatrick \citep{Fitzpatrick99}, 2) updating the SALT2 model \citep{Taylor_2021} to include the modern calibrations \citep{Fragilistic,rykoff2023dark}, and 3) updating from a color-independent SNe intrinsic-scatter model \citep{Brout_2021,Popovic:2021yuo} to a color and dust-dependent physically-motivated model. 
These updates were introduced to improve treatment of the largest sources of systematics in SNe cosmology \cite[Milky Way extinction, light-curve calibration and fitting and SNe Ia intrinsic scatter modelling, ][]{brout2022pantheon+,vincenzi24,rubin2023union} and constitute an objective improvement to previous SNe analyses. Uncertainty in the updates themselves remain accounted for in the reported systematic uncertainties.

The BAO-derived value of $\Lambda$CDM \(\Omega_m\) from DESI Y1 \citep{desiy1} is slightly lower than that obtained from Planck CMB \citep{Planck2018} and SNe \citep{DESY5,brout2022pantheon+,rubin2023union} observations. 
The DESI Year 1 BAO analysis includes thorough evaluations of theoretical modeling uncertainties (estimated to be at most 0.1\% to 0.2\% of the BAO parameters), uncertainties due to the galaxy-halo connection (less than 0.2\%), and observational systematic effects, which are found to be negligible. 
These updates were introduced to improve the treatment of the largest sources of systematics in BAO cosmology and represent objective improvements over previous analyses. Since the systematic contributions are significantly smaller than the statistical uncertainties, unaccounted-for systematics remain unlikely as they would have to be an order of magnitude larger than the current best known systematics to be the cause of the slightly low \(\Omega_m\) inferred from BAO. While the observed discrepancy between BAO and CMB is very small $\sim 1.2\sigma$, it could simply be due to statistical fluctuations or sample/cosmic variance.

An interesting outcome of this work is that we find that for the BAO mock datasets in \( w_0w_a \)CDM background cosmologies, when analyzed under the \(\Lambda\)CDM model, the BAO data exhibits a preference for a slightly higher value of the Hubble constant $\Delta H_0 = $ 1.2km/s/Mpc than predicted by the CMB. 
Inconsistencies between CMB and BAO predicted values for H0 are expected if the modeled late time cosmology is different from the true underlying physics. 
This tendency of BAO data to favor a higher \( H_0 \) qualitatively aligns with the findings from the DESI Year 1 results \citep{desiy1}, where BAO measurements indicate a Hubble constant around \( \sim 67.62\) km/s/Mpc, in contrast to the lower value of \( \sim 66.6 \) km/s/Mpc derived from the mock Planck CMB data \citep{Planck2018}. 

Cumulatively, our results do support the possibility that an underlying dynamical dark energy model could be responsible for the observed discrepancies. This is consistent with the findings of \cite{berghaus2024quantifying}, who showed that scalar field models with non-negligible kinetic energy can alleviate tensions between SNe and BAO datasets. \cite{tada2024quintessential} further explored quintessential dark energy models in light of DESI observations, suggesting that such models can provide better fits to the data. In addition, we have tested the robustness of these conclusions and checked for prior-volume effects using a profile-likelihood analysis (see Appendix~\ref{profilelikelihood}), confirming that they hold even when exploring potential non-Gaussian behaviors in the $w_0w_a$ parameter space.

However, it is important to acknowledge the limitations of our study. Our simulations used idealized datasets without observational noise or scatter, focusing on the impact of cosmological models without the complications introduced by statistical fluctuations. While this simplification allows us to isolate and clearly demonstrate how discrepancies in \(\Omega_m\) and $H_0$ are related to the signal of dynamical dark energy, it does not capture the full complexity of real observational data. In practice, datasets are subject to various sources of uncertainty, including measurement errors, selection effects, calibration issues, and physics systematics, which can influence parameter estimation. Moreover, for the CMB analysis, we relied on compressed CMB parameters - the shift parameter \( R \) and the acoustic scale \( l_A \), with uncertainties derived from Planck 2018 results under the \(\Lambda\)CDM model. While \( R \) and \( l_A \) are designed to be model-independent compressions relevant for dark energy studies, their uncertainties can be model-dependent, especially when extending beyond \(\Lambda\)CDM to models like \( w_0w_a \)CDM. 
For the BAO and CMB analyses, due to concerns with the minimization procedure, we fixed \( \Omega_b h^2 \) to its central value from BBN \citep{BBN}, complicating its uncertainty and the potential Gaussianity assumption in the Hessian matrix maximization analysis. 
One caveat is that simplification may not accurately reflect the true uncertainties in the BAO and CMB constraints, and we do take a Gaussian prior into account for our MCMC analysis. 

In conclusion, the observed discrepancies in \(\Omega_m\) between SNe, BAO, and CMB datasets, and in $H_0$ between BAO and CMB can be explained simultaneously by a dynamical dark energy model. This underscores the necessity for meticulous analysis methods and the careful consideration of systematic uncertainties in cosmological research. 
As upcoming analyses (DES Y6 weak lensing and galaxy clustering, DESI Y3/Y5) and future surveys like the Vera C. Rubin Observatory's LSST and the Euclid mission provide more precise data about background expansion and structure formation, we will soon learn the extent of the significance of the $w_0w_a$ signal and if the growth of structure is also compatible with dynamical dark energy.

\section*{Acknowledgments}

We thank Licia Verde, Sesh Nadathur, Adam Riess, Daniel Scolnic, Dragan Huterer and Tamara Davis for their commentary. We also thank Hongwan Liu, Martin Schmaltz and Nickolas Kokron for extended useful conversations. 

We thank the Templeton Foundation for directly supporting this research. 

All simulations and calculations were performed using Python and several scientific computing packages. The \texttt{Astropy} package \cite{astropy} was used for cosmological computations, such as distance calculations and model evaluations. Numerical computations were carried out using \texttt{NumPy} \citep{numpy2020}, and \texttt{SciPy} \citep{scipy2020} was used for integration and optimization. The Markov Chain Monte Carlo (MCMC) analyses utilized the \texttt{emcee} package \citep{emcee2013}. We use \texttt{matplotlib}, \texttt{corner} \citep{foreman2016corner} and \texttt{Chainconsumer} \citep{ChainConsumer} for plotting.

\appendix
\section{\label{sec:forward} False signal from discrepancies purely in \(\Lambda\)CDM?}

\begin{figure*}
    \centering
    \includegraphics[width=.75\textwidth]{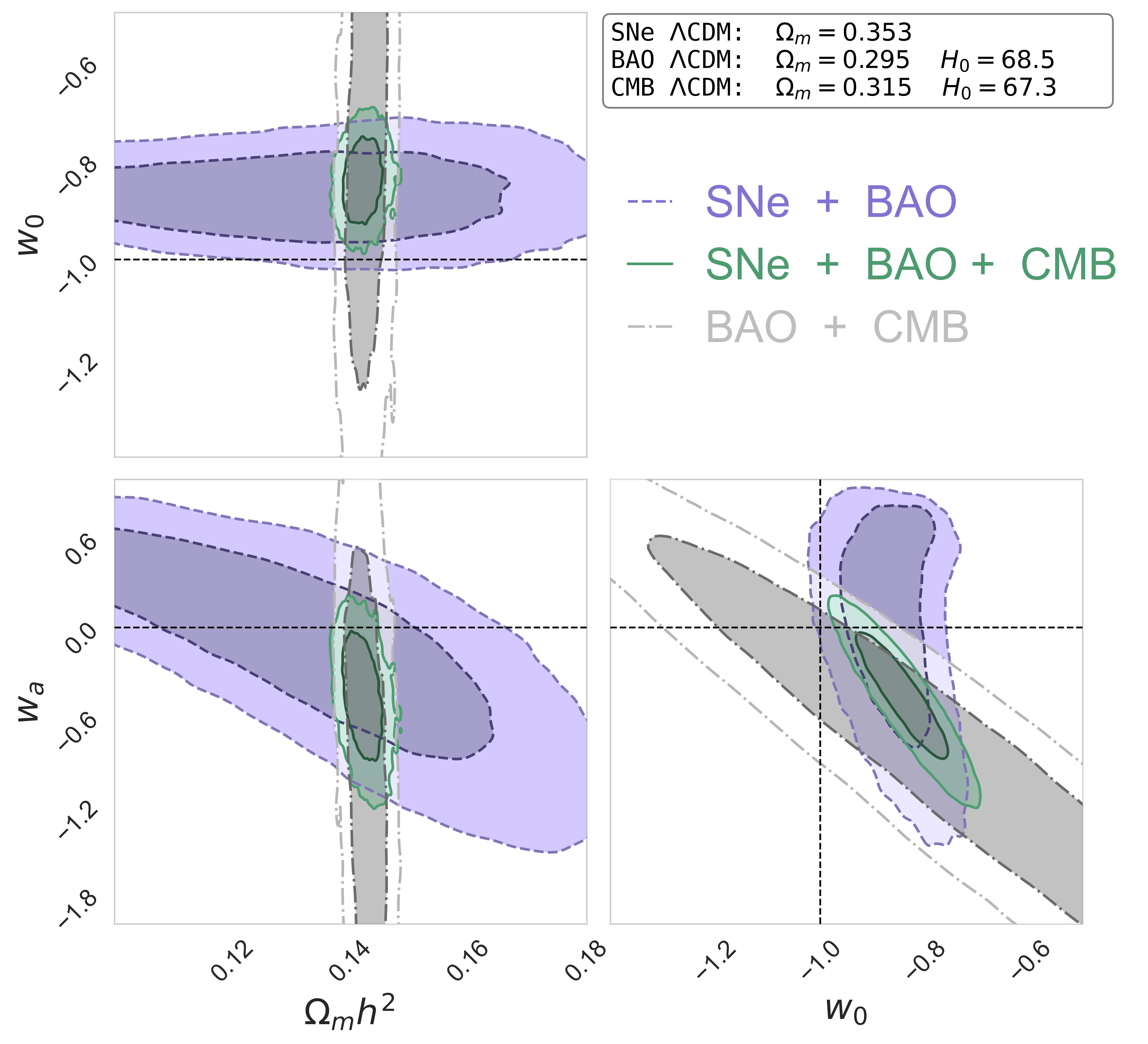}
    \caption{MCMC results testing the \(w_0w_a\)CDM model using mock data combined from different probe-like datasets generated under the \(\Lambda\)CDM model but with different \(\Omega_m\) and \(H_0\). The black dashed lines indicate the \(\Lambda\)CDM values (\(w_0 = -1\), \(w_a = 0\)). The \textbf{purple contour} represents the constraints from mock SNe (\(\Omega_m = 0.353\)) and BAO (\(\Omega_m = 0.295\), \(H_0 = 68.5 \ \text{km/s/Mpc}\)) data. The \textbf{green contour} adds the CMB (\(\Omega_m = 0.315\), \(H_0 = 67.3 \ \text{km/s/Mpc}\)) mock data to the previous combination. The \textbf{grey contour} shows the constraints from BAO and CMB mock data. This figure illustrates how discrepancies in the \(\Lambda\)CDM model's \(\Omega_m\) and \(H_0\) among different datasets can lead to signals of dynamical dark energy when analyzed under the \(w_0w_a\)CDM model.}
    \label{fig:diffom} 
\end{figure*}

If each dataset is consistent with \(\Lambda\)CDM alone, but they are inconsistent with each other (in \(\Omega_m\) and \(H_0\)), could combining them in a joint analysis lead to a false detection of evolving dark energy?  Such \(\Omega_m\) and \(H_0\) discrepancies could be due to fluctuations in the data (either statistical or systematic) that result in the data best matching different central values. 

To address this question to first order, we explicitly simulate each dataset in a \(\Lambda\)CDM cosmology, but with different \(\Omega_m\) and \(H_0\) to mimic either a statistical or systematic fluctuation. It is important to note that this methodology oversimplifies the impact of both statistical fluctuations and systematic errors within one probe and we leave a more realistic analysis for future work. Following Section \ref{sec:sims}, we simulate SNe, BAO and CMB datasets under the \(\Lambda\)CDM model but with different values of \(\Omega_m\) for each probe, consistent with what is observed in real data for each probe individually \citep{desiy1,DESY5,Planck2018}. Similarly, we simultaneously simulate CMB and BAO with different values of \(H_0\). We note that because the SNe likelihood includes absolute magnitude $M$ as an unconstrained variable, the results in this section are insensitive to the value of \(H_0\) chosen for the SNe mocks. The explicit cosmologies simulated for each probe are:
\[
\begin{aligned}
\textbf{SNe Mock \phantom{-.}:} &\; \text{}\Lambda\text{CDM, } \Omega_m = 0.353 \\[5pt]
\textbf{BAO Mock \phantom{.}:} &\; \text{}\Lambda\text{CDM, } \Omega_m = 0.295,~H_0 = 68.5 \ \text{km/s/Mpc} \\[5pt]
\textbf{CMB Mock :} &\; \text{}\Lambda\text{CDM, } \Omega_m = 0.315,~H_0 = 67.3 \ \text{km/s/Mpc}
\end{aligned}
\]

We then perform an MCMC analysis on the combined datasets with log likelihood 
\begin{equation}
    \ln \mathcal{L}_{combined}=\ln \mathcal{L}_{\rm SNe}+\ln \mathcal{L}_{\rm BAO}+\ln \mathcal{L}_{\rm CMB} \,,
\end{equation} 
where the likelihood of each probe is calculated from Eq.~\eqref{eq:snelikelihood}, Eq.~\eqref{eq:baolikelihood} and Eq.~\eqref{eq:cmblikelihood}. We then fit them jointly to a \(w_0w_a\)CDM model to see if the discrepancies in \(\Omega_m\) lead to apparent signals of evolving dark energy and derive constraints on \( \Omega_m \), \(H_0 \), \( w_0 \), and \( w_a \) from the combination of simulated datasets.

\subsection{\label{forward_results}Results From \(\Lambda\)CDM $\Omega_m$ Discrepancies}

Analyzing the combined simulated datasets generated under \(\Lambda\)CDM and fitted under the \(w_0w_a\)CDM model, we find the following constraints when combining only SNe (\(\Omega_m=0.353\)) and BAO (\(\Omega_m=0.295\), \(H_0=68.5 \ \text{km/s/Mpc}\)):
\[
w_0 = -0.856^{+0.054}_{-0.066}, \quad w_a = 0.16^{+0.57}_{-0.50} \,.
\]
These results are shown by the \textbf{purple contour} in Fig.~\ref{fig:diffom}. We observe that \(w_a\) is consistent with zero within uncertainties and \(w_0 \approx -0.86\), indicating no significant deviation from the \(\Lambda\)CDM model.

Next, we analyze the combination of BAO (\(\Omega_m=0.295\), \(H_0=68.5 \ \text{km/s/Mpc}\)) and CMB (\(\Omega_m=0.315\), \(H_0=67.3 \ \text{km/s/Mpc}\)) mock datasets and obtain:
\[
w_0 = -0.70^{+0.37}_{-0.40}, \quad w_a = -0.66^{+0.79}_{-1.38} \,.
\]
These results are represented by the \textbf{grey contour} in Fig.~\ref{fig:diffom}. The central value of \(w_0 = -0.70\) is less negative than the \(\Lambda\)CDM value of \(-1\), but consistent within uncertainties with \(\Lambda\)CDM. Similarly, the central value \(w_a = -0.66\) suggests a mild preference for evolving dark energy, but the large uncertainties make it consistent with zero.

When all the mock datasets are combined—SNe (\(\Omega_m=0.353\)), BAO (\(\Omega_m=0.295\), \(H_0=68.5 \ \text{km/s/Mpc}\)), and CMB (\(\Omega_m=0.315\), \(H_0=67.3 \ \text{km/s/Mpc}\))—we find:
\[
w_0 = -0.851^{+0.065}_{-0.050}, \quad w_a = -0.44^{+0.27}_{-0.30} \,.
\]
These results are shown by the \textbf{green contour} in Fig.~\ref{fig:diffom}. The preferences for \(w_a < 0\) and a more positive \(w_0 \approx -0.85\) become more pronounced, reaching levels of approximately \(2.6\sigma\) for \(w_0\) and \(1.5\sigma\) for \(w_a\). These values suggest that discrepancies in \(\Omega_m\) and \(H_0\) in the \(\Lambda\)CDM model between different datasets can lead to apparent deviations from \(\Lambda\)CDM when analyzed under the \(w_0w_a\)CDM framework, even though each dataset was originally simulated in a \(\Lambda\)CDM universe.

In contrast, when only \(\Omega_m\) discrepancies are present (and \(H_0\) is consistent across datasets), combining all mock datasets gives results that are consistent with \(\Lambda\)CDM values of \(w_0\) and \(w_a\) within \(1\sigma\). This suggests that in order to mimic a dynamical dark energy signal in a universe that is truly \(\Lambda\)CDM, systematic effects need to produce discrepancies in both \(\Omega_m\) and \(H_0\) among different datasets. Such combined discrepancies can lead to apparent preferences for evolving dark energy when analyzing the data under the \(w_0w_a\)CDM framework.

Comparing these results to those observed in the DESI Year 1 combined-probes analysis \citep{desiy1} indicates that discrepancies in \(\Omega_m\) and \(H_0\) can induce apparent deviations from \(\Lambda\)CDM, and the significance of these deviations depends on the magnitude of the discrepancies and the precision of the datasets involved.

\section{Grid-Based Chain-independent Exploration}

So far, we have demonstrated that viable \(w_0w_a\)CDM cosmologies, as constrained by DES24, tend to result in different \(\Omega_m\) when constrained in \(\Lambda\)CDM, similar to those observed in real data. In this section, we aim to present a uniform approach to determining the set of \(w_0w_a\)CDM cosmologies that produce \(\Lambda\)CDM \(\Omega_m\) differences among probes comparable to those observed in read data. By definition, this set will encompass parameter volumes as large as, or larger than, those supported by the DES24 constraints.

We perform a grid-based exploration of the \(w_0w_a\) parameter space.  This approach is independent of specific MCMC chains and allows us to identify regions where the observed discrepancies in \(\Omega_m\) can be reproduced.
We construct a grid over a range of $\{w_0 , w_a , \Omega_m\}$ values, covering \(w_0 \in [-1.2, 0.1]\), \(w_a \in [-6, 1]\), and \(\Omega_m \in [0.23, 0.4]\). For each grid point \((w_0, w_a, \Omega_m)\), we generate mock datasets for SNe, BAO, and CMB using the methods described in Section~\ref{sec:sims}. 
For each mock dataset, we find the \(\Lambda\)CDM best-fit \(\Omega_m\) (as well as $H_0$) values for each probe individually. We seek to compare these fitted \(\Omega_m\) values to the observed values from real data:
\begin{itemize}
    \item \textbf{DES-SN5YR}: \(\Omega_{m,\text{obs}}^{\text{SNe}} = 0.353 \pm 0.017\) \cite{DESY5} \,,
    \item \textbf{DESI BAO}: \(\Omega_{m,\text{obs}}^{\text{BAO}} = 0.295 \pm 0.015\) \cite{desiy1} \,,
    \item \textbf{Planck 2018 CMB}: \(\Omega_{m,\text{obs}}^{\text{CMB}} = 0.315 \pm 0.007\) \cite{Planck2018} \,.
\end{itemize} 
Quantitatively, we compute the likelihood of the best-fit \(\Omega_m\) values being consistent with the observed values for each probe using the Gaussian probability density function
\begin{equation}
\mathcal{L}^*_{\text{probe}} = \exp\left( -\frac{ ( \Omega_{m,\text{fit}} - \Omega_{m,\text{obs}} )^2 }{ 2\sigma_{\text{obs}}^2 } \right ),
\end{equation}
where \(\Omega_{m,\text{fit}}\) is the best-fit value from the \(\Lambda\)CDM analysis of the mock data, and \(\sigma_{\text{obs}}\) is the observed uncertainty.
The joint likelihood for each grid point was calculated by multiplying the likelihoods from all three probes
\begin{equation}
\mathcal{L}^*_{\text{joint}} = \mathcal{L}^*_{\text{SNe}} \times \mathcal{L}^*_{\text{BAO}} \times \mathcal{L}^*_{\text{CMB}} \,.
\label{eq:l*}
\end{equation}

To visualize the results and compare them with observational data, we employ the \texttt{ChainConsumer} package \citep{ChainConsumer}. We create a dataset containing the grid points \((w_0, w_a, \Omega_m)\) and their corresponding joint likelihoods. The weights were calculated as described by Equation~\ref{eq:l*} and normalized. We also include posterior samples from the combined DES24 \(w_0w_a\)CDM chain \citep{DESY5} for comparison.

\begin{figure}
    \centering
    \includegraphics[width=.75\linewidth]{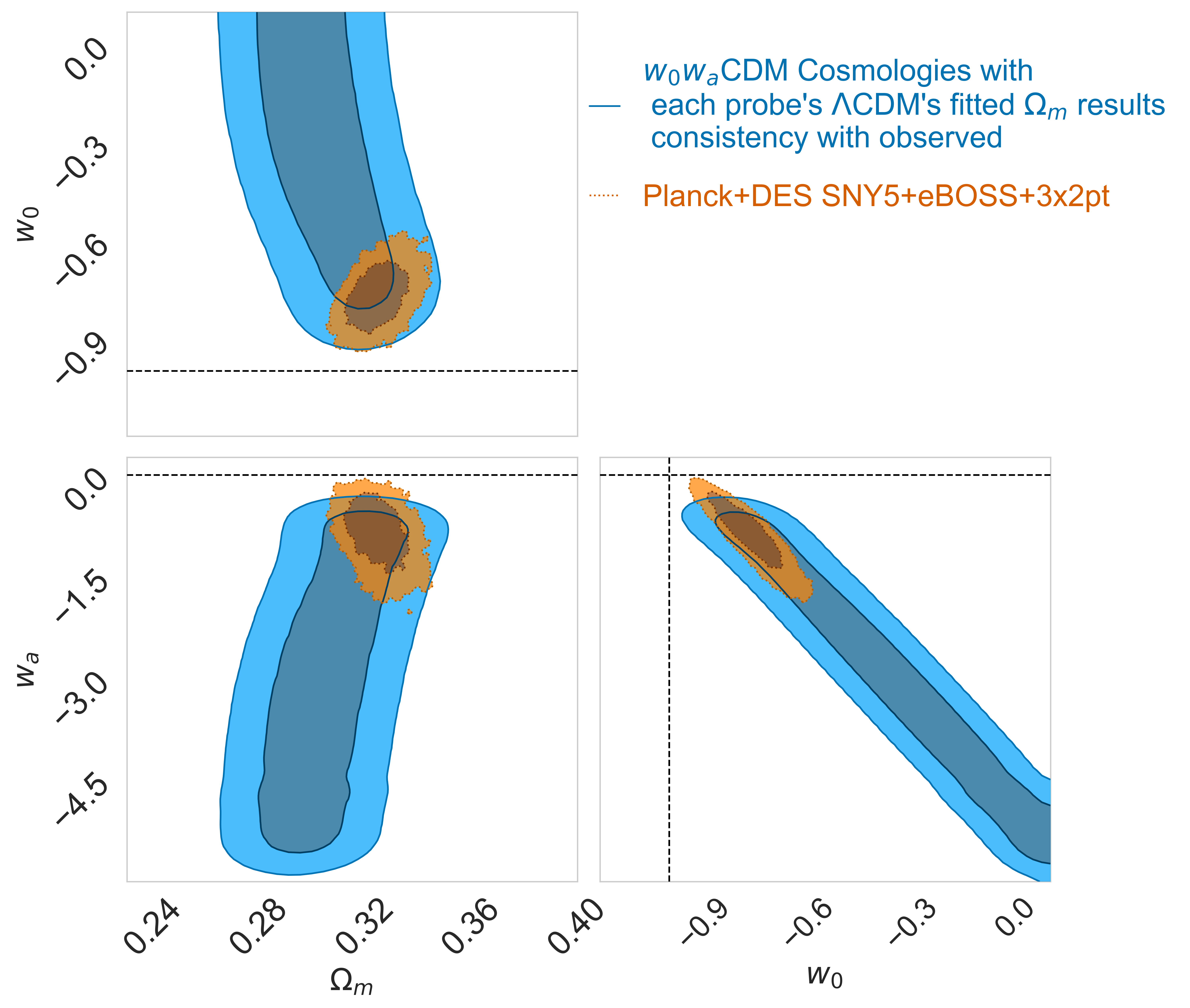}
    \caption{Results of the grid-based exploration of the \(w_0w_a\) parameter space that can generate the observed discrepancies in $\Lambda$CDM $\Omega_m$. Blue contours represent grid points where each probe's fitted $\Lambda$CDM $\Omega_m$s are similar to observed values (which are SNe: $0.353 \pm 0.017$; BAO: $0.295 \pm 0.015$; and CMB: $0.315 \pm 0.007$ \citep{DESY5,desiy1,Planck2018}), with the $1,2\  \sigma$ contours obtained from the likelihood $\mathcal{L}^*_{\text{joint}}$ from Equation~\ref{eq:l*}. The orange contours show the samples from the DES24 chain \citep{DESY5}. Black dashed lines indicate \(\Lambda\)CDM values (\(w_0 = -1\), \(w_a = 0\))}. 
    \label{fig:grid_results}
\end{figure}

The results of our grid-based exploration are summarized in contour plots in Fig.~\ref{fig:grid_results}. The blue contours represent regions where the best-fit \( \Omega_m \) values from each probe's \( \Lambda \)CDM analysis align with the observed \( \Omega_m \) values. Hence, these contours effectively highlight regions in the \( w_0 w_a \) parameter space where each simulated probe individually produces a $\Lambda$CDM \( \Omega_m \) result consistent with its real observational counterpart. The orange contours, on the other hand, show the posterior distribution of \( w_0 \) and \( w_a \) from a combined analysis of DES24 data \citep{DESY5}. The orange DES24 constraints are largely aligned with, and a subset of, our blue grid-based exploration, however we note that there exist many more combinations of cosmological parameters that can reconcile the discrepant \( \Lambda \)CDM  \( \Omega_m \) values (as evidenced by the larger extent of the blue contour).

\section{Impact of Direct Calibration of the Sound Horizon \lowercase{$r_d$}\label{sec:appendix_rd}}

In our main analysis, as stated in Section~\ref{sec:sims}, the BAO data were used with a model-dependent computation of the sound horizon \( r_d \) at the drag epoch, which depends on cosmological parameters such as \( \Omega_m h^2 \) and \( \Omega_b h^2 \). Specifically, we used the scaling relation from Eq. (2.5) in the DESI Y1 paper \citep{desiy1} and BBN in our main analysis. We note that the \( r_d \) constraints we applied are from a Planck \(\Lambda\)CDM fit. In this section, we use mock data simulated in $w_0w_a$CDM cosmologies, then fit them assuming \(\Lambda\)CDM. The \( r_d \) constraints might therefore be less reliable here because we are simulating in $w_0w_a$CDM cosmologies. The direct \( r_d \) calibration was included here to maintain consistency with the \(\Lambda\)CDM fit in the DESI Y1 BAO analysis \citep{desiy1}.

Closely following the methodology used by the DESI collaboration in their Year 1 results \citep{desiy1}, we explore the impact of directly calibrating \( r_d \) using external constraints. Specifically, we adopt a Gaussian prior on \( r_d \) based on CMB measurements
\begin{equation}
    r_d = 147.09 \pm 0.26 \ \text{Mpc} \,.
\end{equation}

This direct calibration effectively decouples \( r_d \) from the late-time cosmological parameters, thereby breaking the degeneracy between \( H_0 \) and \( r_d \) inherent in BAO measurements when \( r_d \) is computed from the cosmological model.

We modify our BAO analysis by directly calibrating \( r_d \). In practice, this means that when computing the BAO observables, we treat \( r_d \) as an external parameter drawn from a Gaussian distribution with the mean and standard deviation specified above.

\begin{figure}
    \centering
    \includegraphics[width=.65\linewidth]{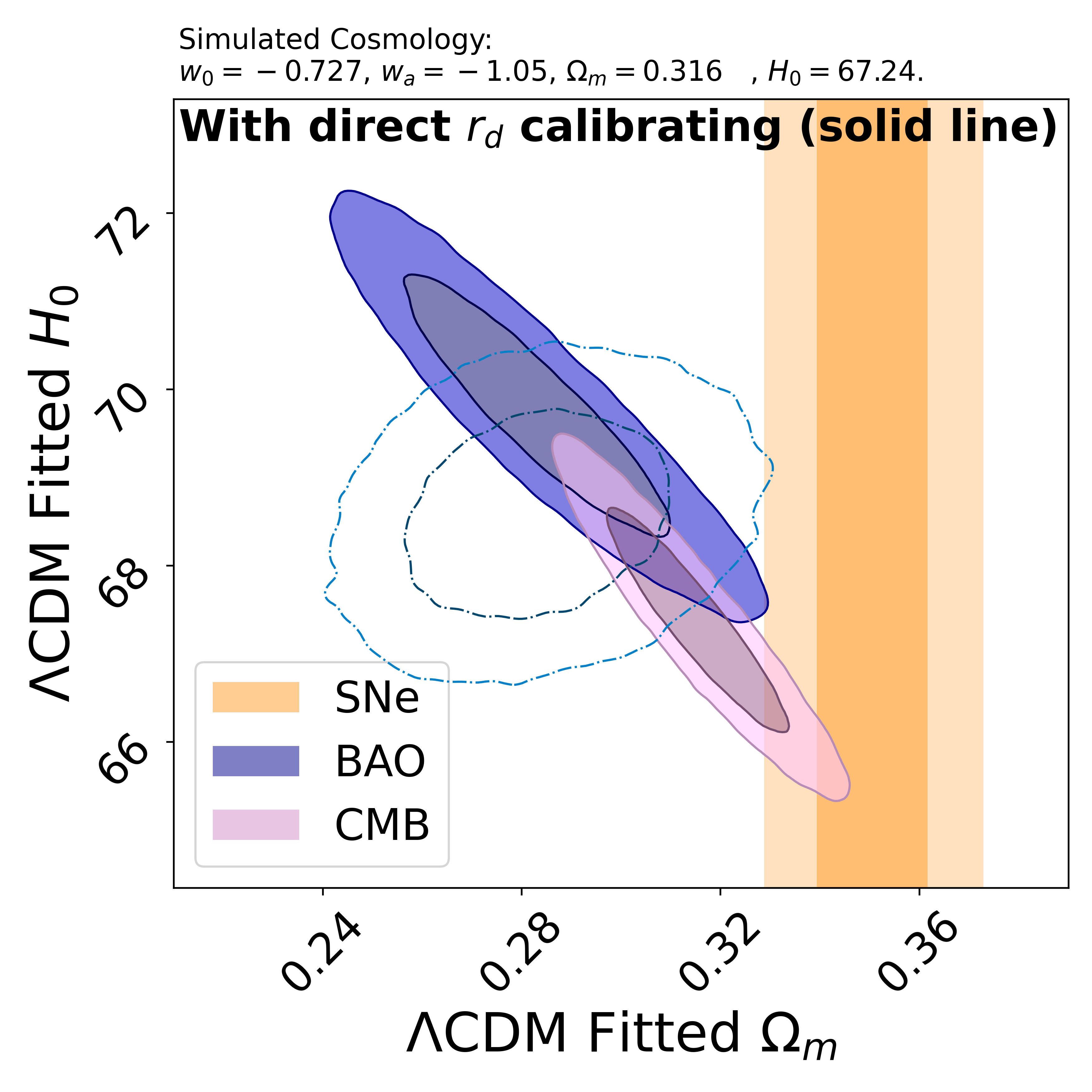}
    \caption{Constraints on \( \Omega_m \) and \( H_0 \) obtained by fitting the \(\Lambda\)CDM model to mock CMB, BAO, and SNe datasets generated in the best-fit \( w_0w_a \)CDM cosmology (\( w_0 = -0.727 \), \( w_a = -1.05 \), \( \Omega_m = 0.316 \), \( H_0 = 67.24\ \text{km/s/Mpc} \)), with direct \( r_d \) calibration. The BAO constraints (dark blue contours) become less circular in the \( \Omega_m \)--\( H_0 \) plane compared to the case without direct \( r_d \) calibration (Fig.~\ref{fig:bestw0wa_to_lcdm}, also shown here as the dash-dot contours).}
    \label{fig:bestw0wa_to_lcdm_desi_gaussian}
\end{figure}

\begin{figure}
    \centering
    \includegraphics[width=.75\linewidth]{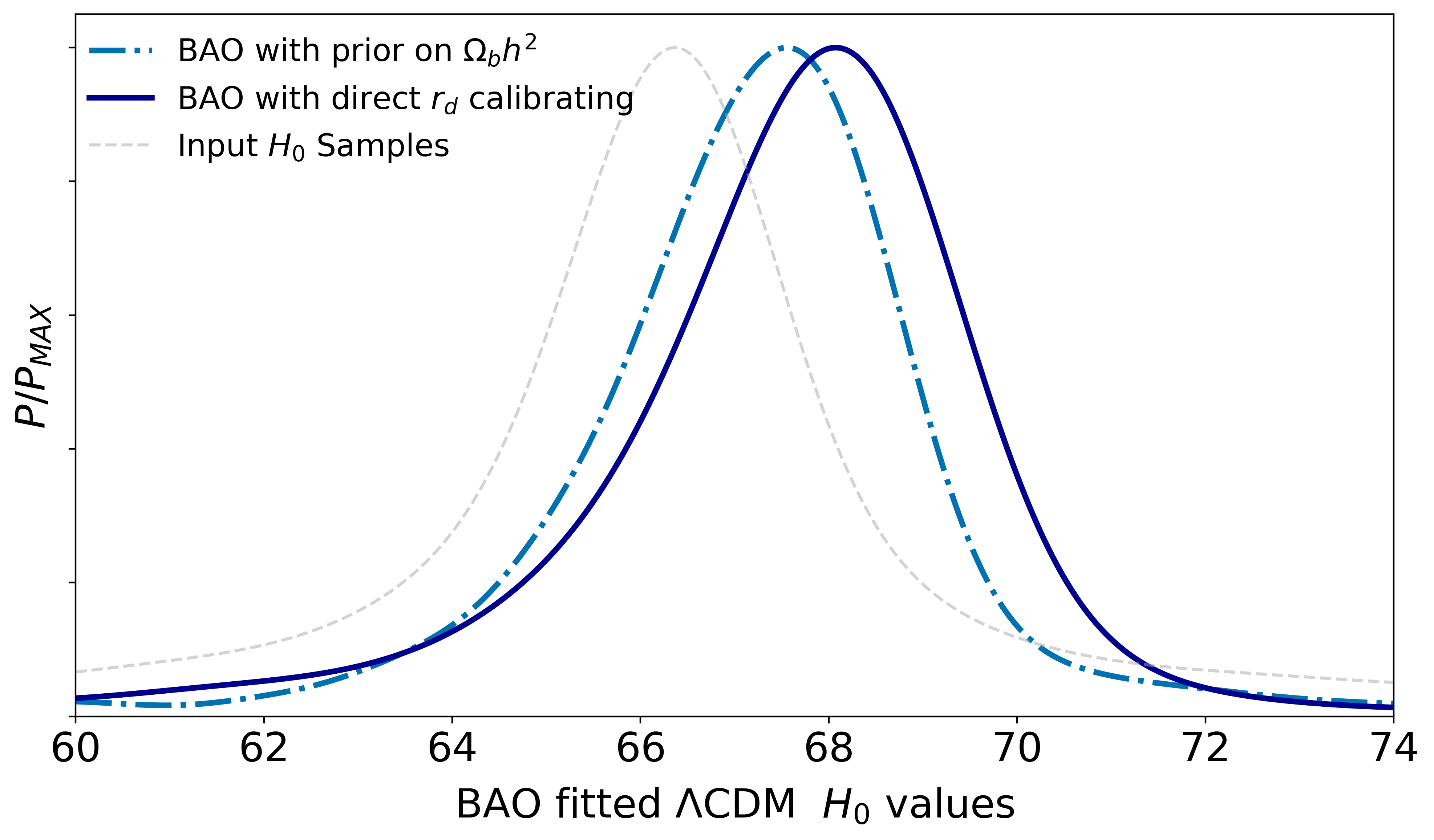}
    \caption{Probability distributions of the fitted \( H_0 \) values (in km/s/Mpc) from BAO under the \(\Lambda\)CDM model, with (blue solid line) and without (blue dash-dot line, same as the one in the main text, right panel of Fig.~\ref{fig:H0_cmb_bao}) direct \( r_d \) calibration. The input \( H_0 \) values from the DES24 \( w_0w_a \)CDM background cosmologies are indicated by the gray dashed line. In both the direct \( r_d \) calibration and using a prior on \(\Omega_b h^2\), the BAO data prefer a higher \( H_0 \) than what was simulated. }
    \label{fig:H0_cmb_bao_rdrag}
\end{figure}

\paragraph{Constraints in the \( \Omega_m \)--\( H_0 \) plane:}
Fig.~\ref{fig:bestw0wa_to_lcdm_desi_gaussian} (direct \( r_d \) calibrating version of Fig.~\ref{fig:bestw0wa_to_lcdm}) shows the constraints on \( \Omega_m \) and \( H_0 \) obtained by fitting the \(\Lambda\)CDM model to mock datasets generated in the DESI Y1 best-fit \( w_0w_a \)CDM cosmology \citep{desiy1}. With the direct \( r_d \) calibration, the BAO constraints become less circular as additional degeneracies open up in the \( \Omega_m \)--\( H_0 \) plane compared to when only \( \Omega_bh^2 \) is provided (blue contours in Fig.~\ref{fig:bestw0wa_to_lcdm} and Fig.~\ref{fig:bestw0wa_to_lcdm_desi_gaussian}).

\paragraph{Hubble Constant Discrepancy:} Despite the direct calibration of \( r_d \), we still observe a discrepancy in the best-fit \( H_0 \) values when the true cosmology is \( w_0w_a \)CDM but data are analyzed under the \(\Lambda\)CDM model. Fig.~\ref{fig:H0_cmb_bao_rdrag} illustrates that the BAO data prefer a higher \( H_0 \), consistent with our findings in the main text in Section~\ref{sec:backward_results} and Fig.~\ref{fig:H0_cmb_bao}. The direct calibration of \( r_d \) has a slightly larger \( H_0 \) than taking a prior on \(\Omega_bh^2\). The difference between the simulated \(w_0w_a\)CDM maximum likelihood $H_0$ and observed direct calibration of \( r_d \) with BAO in \(\Lambda\)CDM maximum likelihood $H_0$ grows to 1.67km/s/Mpc, which is a slightly larger difference than when using the prior on \(\Omega_bh^2\) (1.22km/s/Mpc).This difference in inferred \( H_0 \) for \( r_d \) calibrated and using a prior on \(\Omega_bh^2\) is quantitatively similar to DESI Y1's results in Eqs. (4.3) \& (4.4) in \cite{desiy1}.
\section{Fisher Matrix Analysis}

In this section, we perform a Fisher matrix analysis to quantify how sensitive our cosmological observables are to the parameters \(\Omega_m\) and \(H_0\). This analysis serves as a cross-check of the constraints derived from our mock datasets. Additionally, Fisher formalism provides a theoretical approach that allows for a simplified computation of parameter constraints, demonstrating how different probes—at various redshifts or precision levels—contribute to shaping the parameter space constraints, without relying on mock data.

The Hubble parameter:
\[
E(z) = \sqrt{\Omega_m(1+z)^3 + (1-\Omega_m)(1+z)^{3(1+w_0+w_a)} e^{-\frac{3 w_a z}{1+z}}}.
\]
\subsection{Fisher Matrix Construction}
\subsubsection{SNe Fisher:}
The derivative of SNe data point:
\[
\frac{\partial d_L(z)}{\partial \Omega_m} = (1+z)\frac{c}{H_0} \int_0^z \frac{ -\frac{(1+z')^3 -1}{2E(z')^3} }{E(z')} dz'
\]

The Fisher matrix \(F^{\text{SNe}}\) for SNe data is:
\[
F^{\text{SNe}} = \sum_{n,m=1}^N \frac{\partial \mu(z_n)}{\partial \Omega_m} \left[\mathbf{C_{\text{DES}}}^{-1}\right]_{nm} \frac{\partial \mu(z_m)}{\partial \Omega_m},
\]

Where \(N\) is the number of SNe data points.

While the derivative of the distance modulus \(\mu(z)\) with respect to \( H_0 \) is theoretically meaningful, it is not included in our Fisher matrix analysis because the intrinsic absolute magnitude \( M \) of a supernova is unknown. Without a precise calibration of \( M \), SNe data alone cannot constrain \( H_0 \). The degeneracy between \( M \) and \( H_0 \) in the distance modulus equation:
\[
\mu(z) = 5\log_{10}[d_L(z)] + 25 - M
\]
renders \( H_0 \) effectively unconstrained without additional external calibration. As a result, our Fisher analysis for SNe focuses solely on constraints for \(\Omega_m\), assuming \( H_0 \) remains degenerate with \( M \).

\subsubsection{BAO Fisher:}
The derivative of BAO data point:
\[
\begin{aligned}
\frac{\partial D_V(z)}{\partial \Omega_m} &= D_V(z) \left[ \frac{2}{3} \frac{\frac{\partial I(z)}{\partial \Omega_m}}{I(z)} - \frac{1}{3} \frac{\frac{\partial E(z)}{\partial \Omega_m}}{E(z)} \right],
&\quad
\frac{\partial D_V(z)}{\partial H_0} &= -\frac{D_V(z)}{H_0};
\\
\frac{\partial F_{\text{AP}}(z)}{\partial \Omega_m} &= \frac{\partial E(z)}{\partial \Omega_m} \cdot I(z) + E(z) \cdot \frac{\partial I(z)}{\partial \Omega_m},
&\quad
\frac{\partial F_{\text{AP}}(z)}{\partial H_0} &= 0;
\end{aligned}
\]

where \( I(z) \equiv \int_0^z \frac{dz'}{E(z')} \), and $\frac{\partial I(z)}{\partial \Omega_m} = \int_0^z -\frac{1}{E(z')^2} \frac{\partial E(z')}{\partial \Omega_m} dz'$;

The Fisher matrix \( F^{\text{BAO}}_{ij} \) for BAO data is constructed by summing contributions from both \( D_V(z) \) and \( F_{\text{AP}}(z) \) measurements:
\[
F_{\text{BAO}} = 
\sum_{k=1}^{N_{D_V}} \frac{1}{\sigma_{D_V,k}^2} 
\begin{pmatrix}
\frac{\partial D_V(z_k)}{\partial \Omega_m} \\
\frac{\partial D_V(z_k)}{\partial H_0}
\end{pmatrix}
\begin{pmatrix}
\frac{\partial D_V(z_k)}{\partial \Omega_m} & \frac{\partial D_V(z_k)}{\partial H_0}
\end{pmatrix}
+
\sum_{l=1}^{N_{F_{\text{AP}}}} \frac{1}{\sigma_{F_{\text{AP}},l}^2} 
\begin{pmatrix}
\frac{\partial F_{\text{AP}}(z_l)}{\partial \Omega_m} \\
\frac{\partial F_{\text{AP}}(z_l)}{\partial H_0}
\end{pmatrix}
\begin{pmatrix}
\frac{\partial F_{\text{AP}}(z_l)}{\partial \Omega_m} & \frac{\partial F_{\text{AP}}(z_l)}{\partial H_0}
\end{pmatrix}.
\]

\( N_{D_V} \) and \( N_{F_{\text{AP}}} \) are the number of BAO data points for \( D_V(z) \) and \( F_{\text{AP}}(z) \), respectively, and \( \sigma_{D_V,k} \), \( \sigma_{F_{\text{AP}},l} \) are the uncertainties associated with each measurement.

\subsubsection{CMB Fisher:}

Due to the complexity of the analytical expressions for the derivatives of \( R \) and \( l_A \) with respect to the cosmological parameters, we compute these partial derivatives numerically using finite difference. Detailed analytical derivatives are available in our supplementary material\footnote{Analytical derivatives of our CMB parameters can be found at \url{https://github.com/trivialTZ/DM_DE_Signals_in_SNe_BAO_CMB/blob/main/Fisher_d_cmb.pdf}}.
\[
\mathbf{J}_{\text{CMB}} = \begin{pmatrix}
\frac{\partial R}{\partial \Omega_m} & \frac{\partial R}{\partial H_0} \\
\frac{\partial l_A}{\partial \Omega_m} & \frac{\partial l_A}{\partial H_0}
\end{pmatrix}.
\]

The Fisher matrix \( \mathbf{F}_{\text{CMB}} \) for the CMB data is constructed as:
\[
\mathbf{F}_{\text{CMB}} = \mathbf{J}_{\text{CMB}}^T \mathbf{C_{P18}}^{-1} \mathbf{J}_{\text{CMB}} \,,
\]
Here, \( \mathbf{C_{P18}}^{-1} \) is the inverse of the covariance matrix defined in Equation~\eqref{eq:covariance}.
\subsection{Parameter Estimation}

The maximum likelihood estimates for the cosmological parameters are obtained by minimizing the Fisher matrix formalism for each dataset. Each dataset contributes its own Fisher matrix and Jacobian matrix, which are combined to derive joint constraints on the cosmological parameters \( \Omega_m \) and \( H_0 \).

\textbf{For SNe data}, the parameter shifts are defined as:
\[
\Delta \mu = \mu_{\text{fid}} - \mu_{\Lambda\text{CDM}},
\]
\[
\vec{\Omega_m}_{\text{ML}}^{(\text{SNe})} = \vec{\Omega_m}_{\text{fid}} + \mathbf{F}_{\text{SNe}}^{-1} \mathbf{J}_{\text{SNe}}^T \mathbf{C}_{\text{DES}}^{-1} \Delta \mu \,.
\]

The uncertainties in the parameters are given by the square roots of the diagonal elements of the inverse Fisher matrix:
\[
\sigma_{\Omega_m}^{\text{SNe}} = \sqrt{\mathbf{F}_{\text{SNe}}^{-1}} \,. 
\]

\textbf{For BAO data}, the parameter shifts are calculated as:
\( \Delta \vec{\theta}_{\text{BAO}} \) are computed as:
\[
\Delta \vec{\theta}_{\text{BAO}} = \mathbf{F}_{\text{BAO}}^{-1} 
\begin{pmatrix}
\sum_{k=1}^{N_{D_V}} \frac{ \frac{\partial D_V(z_k)}{\partial \Omega_m} \Delta \text{BAO}_k }{ \sigma_{D_V,k}^2 } + \sum_{l=1}^{N_{F_{\text{AP}}}} \frac{ \frac{\partial F_{\text{AP}}(z_l)}{\partial \Omega_m} \Delta \text{BAO}_l }{ \sigma_{F_{\text{AP}},l}^2 } \\
\sum_{k=1}^{N_{D_V}} \frac{ \frac{\partial D_V(z_k)}{\partial H_0} \Delta \text{BAO}_k }{ \sigma_{D_V,k}^2 }
\end{pmatrix}.
\]
Thus, the maximum likelihood estimates are:
\[
\vec{\theta}_{\text{ML}}^{\ (\text{BAO})} = \vec{\theta}_{\text{fid}} + \Delta \vec{\theta}_{\text{BAO}} \,.
\]
with uncertainties:
\[
\sigma_{\Omega_m}^{\text{BAO}} = \sqrt{\left(\mathbf{F}_{\text{BAO}}^{-1}\right)_{11}}, \quad \sigma_{H_0}^{\text{BAO}} = \sqrt{\left(\mathbf{F}_{\text{BAO}}^{-1}\right)_{22}}.
\]

\textbf{For CMB data}, the parameter shifts are computed as:
\[
\Delta_{\text{CMB}} = \begin{pmatrix}
R_{\Lambda\text{CDM}} - R_{\text{fid}} \\
l_{A,\Lambda\text{CDM}} - l_{A,\text{fid}}
\end{pmatrix},
\]
\[
\vec{\theta}_{\text{ML}}^{(\text{CMB})} = \vec{\theta}_{\text{fid}} + \mathbf{F}_{\text{CMB}}^{-1} \mathbf{J}_{\text{CMB}}^T \mathbf{C}_{\text{P18}}^{-1} \Delta_{\text{CMB}} \,.
\]
The uncertainties in the parameters are given by the square roots of the diagonal elements of the inverse Fisher matrix:
\[
\sigma_{\Omega_m}^{\text{CMB}} = \sqrt{\left(\mathbf{F}_{\text{CMB}}^{-1}\right)_{11}}, \quad \sigma_{H_0}^{\text{CMB}} = \sqrt{\left(\mathbf{F}_{\text{CMB}}^{-1}\right)_{22}}.
\]

\begin{figure}
    \centering
    \includegraphics[width=0.65\linewidth]{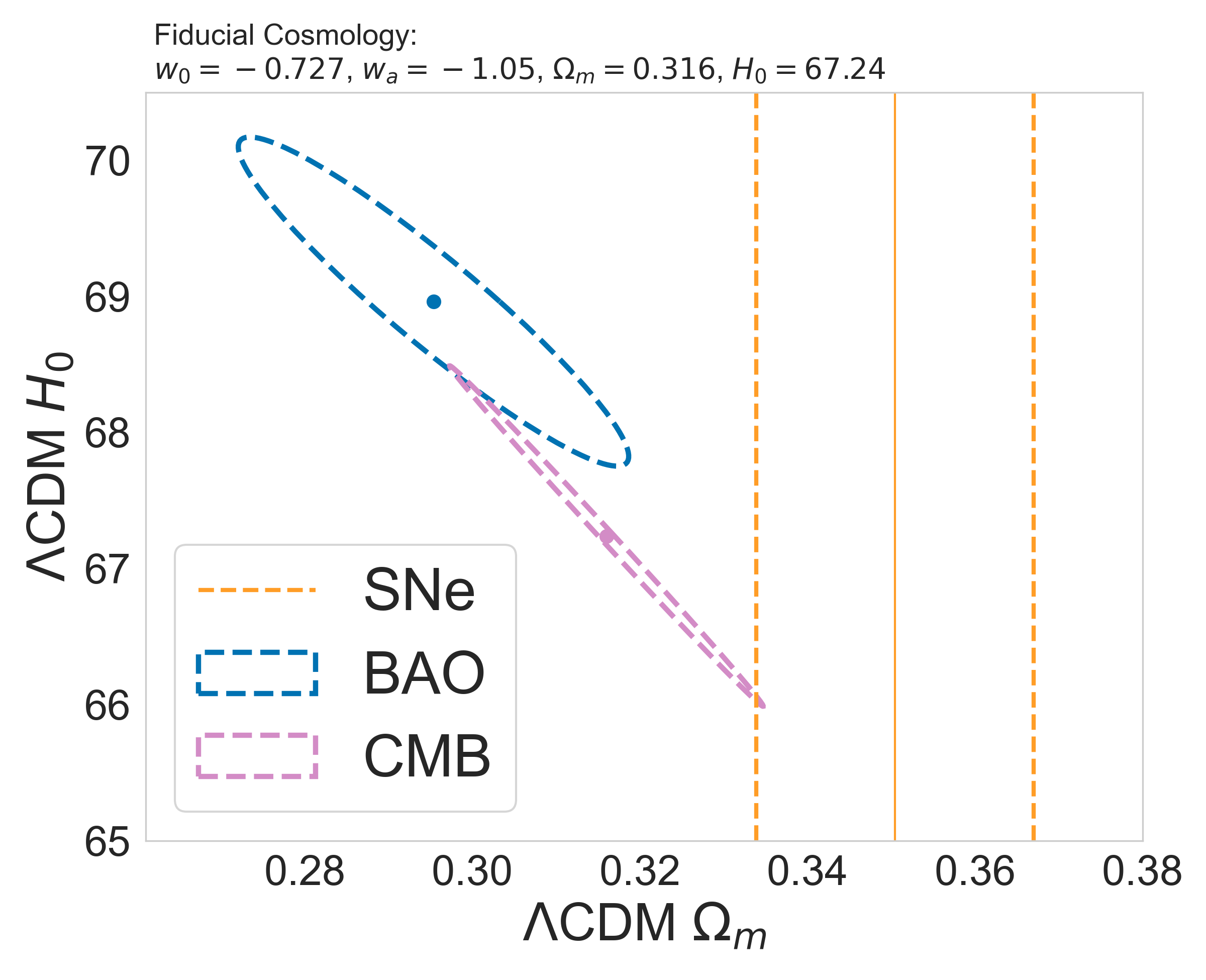}
    \caption{
    Fisher matrix constraints on \(\Omega_m\) and \(H_0\) under \(\Lambda\)CDM framework derived from mock datasets under the best-fit \(w_0w_a\)CDM cosmology from DESI+DES-SN+Planck. The SNe constraints are shown as vertical dashed lines with a solid central line (yellow), BAO constraints are represented by blue ellipses, and CMB constraints are shown as pink ellipses, The solid point/line is the Maximum likelihood point calculated from Fisher. The fiducial cosmology \( \{w_0 = -0.727, w_a = -1.05, \Omega_m = 0.316, H_0 = 67.24\} \).}
    \label{fig:fisher_constraints}
\end{figure}

We present the constraints on \(\Omega_m\) and \(H_0\) derived from the Fisher matrix analysis applied to mock CMB, BAO, and SNe datasets generated under the best-fit \( w_0w_a \)CDM cosmology from DESI+DES-SN+Planck \citep{desiy1}: \( \{w_0 = -0.727, w_a = -1.05, \Omega_m = 0.316, H_0 = 67.24 \} \). The results are shown in Fig.~\ref{fig:fisher_constraints}, where we plot the \( 68\% \) confidence regions for each probe individually and their respective maximum likelihood estimates.

Additionally, the numerical results of our Fisher matrix analysis are summarized below:
\begin{itemize}
    \item \textbf{SNe Results:}
    \[
    \Omega_m^{\text{ML}} = 0.350 \pm 0.017
    \]

    \item \textbf{BAO Results:}
    \[
    \Omega_m^{\text{ML}} = 0.295 \pm 0.015, \quad H_0^{\text{ML}} = 68.96 \pm 0.80 \,
    \]

    \item \textbf{CMB Results:}
    \[
    \Omega_m^{\text{ML}} = 0.314 \pm 0.012, \quad H_0^{\text{ML}} = 67.40 \pm 0.83 \,
    \]
\end{itemize}

The constraints presented in Fig.~\ref{fig:fisher_constraints} can be compared with those in Figs.~\ref{fig:bestw0wa_to_lcdm} and \ref{fig:bestw0wa_to_lcdm_desi_gaussian}. In Fig.~\ref{fig:bestw0wa_to_lcdm_desi_gaussian}, the BAO constraints are recalculated under the assumption of direct calibration of \( r_d \), which aligns with our Fisher matrix results. Specifically, in our Fisher matrix analysis, the BAO calculations treat \( r_d \) as a constant, making the constraints more elongated and less circular in the \(\Omega_m\)--\(H_0\) plane compared to Fig.~\ref{fig:bestw0wa_to_lcdm}. Our Fisher matrix results align closely with the findings in Fig.~\ref{fig:bestw0wa_to_lcdm_desi_gaussian}, where direct \( r_d \) calibration is applied.

\section{\label{sec:null} Null Tests}

To validate our analysis methods, we performed a null test wherein we generated mock datasets under a consistent \(\Lambda\)CDM cosmology and analyzed them using the \(w_0w_a\)CDM cosmological model. This test aims to confirm that our pipeline correctly recovers the input cosmological parameters when there are no discrepancies among the probes and that it does not falsely indicate a preference for evolving dark energy.

We simulated SNe, BAO, and CMB datasets using the same \(\Lambda\)CDM cosmology with \(H_0 = 70\) km/s/Mpc and \(\Omega_m = 0.30\). This choice ensures that all probes are consistent with each other and with the cosmological model used in the analysis.

We then performed the MCMC analysis on the combined mock datasets as in Section~\ref{sec:backward_method} and Appendix~\ref{sec:forward}, fitting them under the \(w_0w_a\)CDM model with additional free parameters. The likelihood function used is the same as in Section~\ref{sec:sims}.

\begin{figure*}
    \centering
    \includegraphics[width=0.95\textwidth]{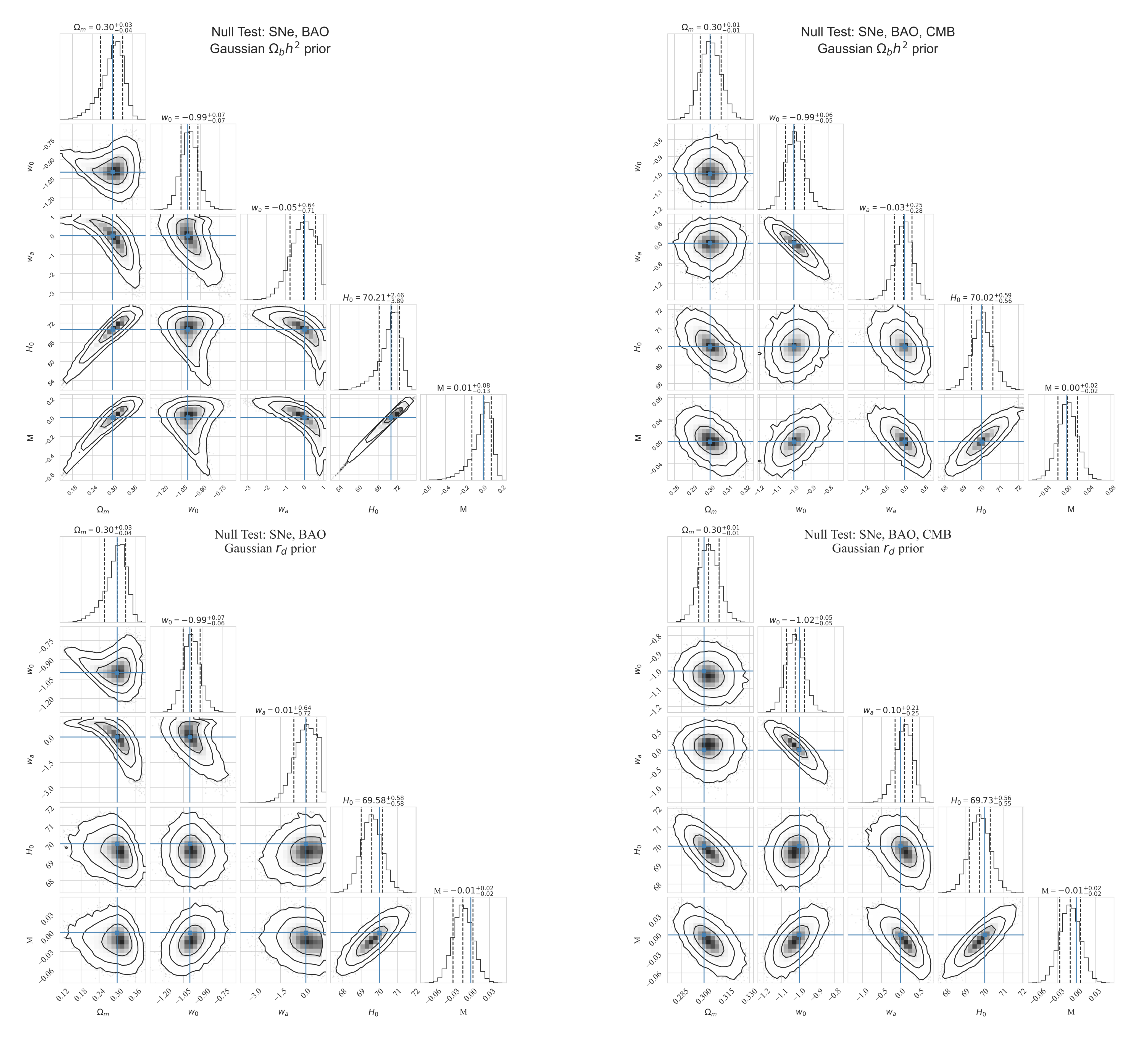}
    \caption{\textbf{Null test}: Posterior distributions of \(\Omega_m\), \(w_0\), \(w_a\), and \(H_0\) from the MCMC analysis of the combined datasets when the input is simulated under the same \(\Lambda\)CDM cosmology. The top two panels (\textbf{Left:} SNe, BAO; \textbf{Right:} SNe, BAO, CMB) adopt \(r_d\) derived using Eq. (2.5) from the DESI Y1 paper which has standard early-time physics assumptions, taking a Gaussian prior on \(\Omega_bh^2=0.02218 \pm 0.00055\), the same as in Eq. (2.9) in \cite{desiy1}. The bottom two panels (\textbf{Left:} SNe, BAO; \textbf{Right:} SNe, BAO, CMB) use a Gaussian prior on the sound horizon \(r_d = 147.09 \pm 0.26\) Mpc.  The input values \(\Omega_m = 0.30\), \(H_0 = 70\) km/s/Mpc are indicated by blue lines, and the contours represent the 68\%, 95\%, and 99.7\% confidence levels.
    All posteriors recover the input cosmology within $1\sigma$. }
    \label{fig:null_test_corner_sne_bao_cmb}
\end{figure*}

The best-fit parameters and their 68\% confidence intervals for the combined datasets are ($H_0$ in $\text{km/s/Mpc}$ unit):

\textbf{Top panels of Fig.~\ref{fig:null_test_corner_sne_bao_cmb}: Gaussian prior on \(\Omega_bh^2\)}:
\[
\begin{aligned}
    \text{SNe + BAO:} & \quad \Omega_m = 0.30^{+0.03}_{-0.04} \quad w_0 = -0.99^{+0.07}_{-0.07} \\
    & \quad w_a = -0.05^{+0.64}_{-0.71} \quad H_0 = 70.21^{+2.46}_{-3.89} \\
    \text{SNe + BAO + CMB:} & \quad \Omega_m = 0.30^{+0.01}_{-0.01} \quad w_0 = -0.99^{+0.06}_{-0.05} \\
    & \quad w_a = -0.03^{+0.25}_{-0.28} \quad H_0 = 70.02^{+0.59}_{-0.56} \\
\end{aligned}
\]

\textbf{Bottom panels of Fig.~\ref{fig:null_test_corner_sne_bao_cmb}: Gaussian prior on \(r_d\)}:
\[
\begin{aligned}
    \text{SNe + BAO:} & \quad \Omega_m = 0.30^{+0.03}_{-0.04} \quad w_0 = -0.99^{+0.07}_{-0.07} \\
    & \quad w_a = 0.01^{+0.64}_{-0.72} \quad H_0 = 69.58^{+0.58}_{-0.58} \\
    \text{SNe + BAO + CMB:} & \quad \Omega_m = 0.30^{+0.01}_{-0.01} \quad w_0 = -1.02^{+0.05}_{-0.05} \\
    & \quad w_a = 0.10^{+0.25}_{-0.21} \quad H_0 = 69.73^{+0.56}_{-0.55} \\
\end{aligned}
\]

The results of the MCMC analysis are shown in Fig.~\ref{fig:null_test_corner_sne_bao_cmb}, which displays the posterior distributions of \(\Omega_m\), \(w_0\), \(w_a\), and \(H_0\). The blue lines indicate the input cosmological parameters used to generate the mock data (\(\Omega_m = 0.30\), \(w_0 = -1\), \(w_a = 0\), \(H_0 = 70\) km/s/Mpc).
They all recover the \(\Lambda\)CDM results well.

\section{Profile Likelihood Tests}
\label{profilelikelihood}

While the main body of this paper focuses on Bayesian and Fisher/Hessian-based approaches, here we present additional \textit{profile-likelihood} tests to validate the robustness of our conclusions to prior-volume effects. Profile likelihoods provide a complementary frequentist framework with a different treatment of nuisance parameters than standard Bayesian marginalization. They can be particularly illuminating if one suspects prior-volume effects or non-Gaussian shapes in the likelihood \citep{herold2024profile_likelihood}.

We construct the profile likelihood for each cosmological parameter using Procoli \citep{Karwal:2024qpt} by fixing that parameter of interest to a grid of values while maximizing (profiling) over all remaining parameters in the likelihood. We then compare these profile-likelihood curves to our MCMC posteriors for the same simulated datasets (described in Section~\ref{sec:sims}). 

\begin{figure}
    \centering
    \includegraphics[width=0.8\linewidth]{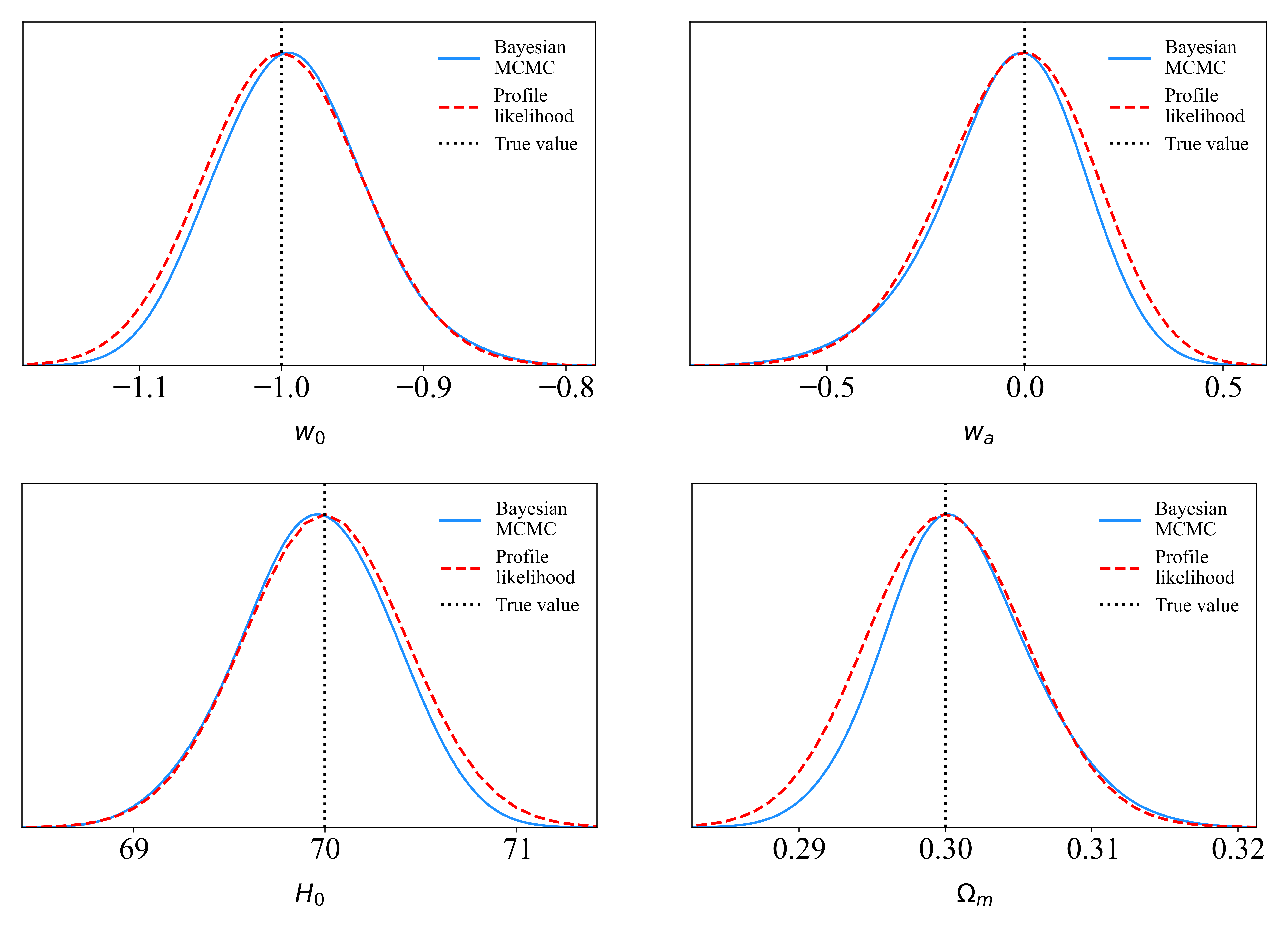}
    \caption{Illustrative profile-likelihood curves (dashed) for cosmological parameters under the $w_0w_a$CDM model, compared with the 1D posterior from our MCMC fit (solid). 
    These were generated for mock CMB, BAO and SNe data in a $\Lambda$CDM universe, marked by the dotted vertical lines.  }
    \label{fig:profilelike_example}
\end{figure}

Figure~\ref{fig:profilelike_example} shows profile-likelihood curves for $w_0$, $w_a$, $\Omega_m$ and $H_0$ for the $w_0w_a$CDM cosmology. We find that the best-fit points and widths of the 1D profile-likelihood constraints closely match the posteriors from our MCMC runs, showing the consistency of our fits.

Overall, the profile-likelihood analysis confirms that for a universe that is truly $\Lambda$CDM and consistent across SNe, BAO and CMB datasets, jointly analyzing it under a $w_0w_a$ cosmology does not introduce any biases in the posteriors of $w_0$, $w_a$, $\Omega_m$ and $H_0$ due to prior-volume effects.

\bibliography{sample631}{}
\bibliographystyle{aasjournal}

\end{document}